\documentclass[12pt]{article}

\pdfoutput=1
\usepackage{graphicx}
\usepackage{amssymb}
\usepackage{amsmath}
\usepackage{color}
\usepackage{subfigure}
\usepackage[colorlinks=true,linkcolor=blue,citecolor=blue]{hyperref}
\begin{document}

\newcommand{\be}{\begin{equation}}
\newcommand{\ee}{\end{equation}}

%\vskip 1cm

\begin{titlepage}

\begin{flushright}
ICRR-Report-646-2012-35  \\
IPMU 13-0021\\
UT-13-02
\end{flushright}

\vskip 2cm

\begin{center}

{\large \bf
Heavy gravitino in hybrid inflation
}

\vspace{1cm}

{Masahiro Kawasaki$^{(a, b)}$, Naoya Kitajima$^{(a)}$, Kazunori Nakayama$^{(c,b)}$\\
and Tsutomu T. Yanagida$^{(b)}$}

\vskip 1.0cm

{\it
$^a$Institute for Cosmic Ray Research,
     University of Tokyo, Kashiwa, Chiba 277-8582, Japan\\
$^b$Kavli Institute for the Physics and Mathematics of the Universe, 
     University of Tokyo, Kashiwa, Chiba 277-8568, Japan\\
$^c$Department of Physics, University of Tokyo, Bunkyo-ku, Tokyo 113-0033, Japan
}

\vskip 1.0cm

\begin{abstract}
	It is known that supersymmetric hybrid inflation model may require severe 
	tunings on the initial condition for large gravitino mass of order 100~-~1000~TeV 
	due to the constant term in the superpotential.
	We propose a modified hybrid inflation model, where the constant term is 
	suppressed during inflation and generated after inflation by replacing 
	a constant term with dynamical field.
	In this modified model, successful inflation consistent with large gravitino mass 
	takes place without severe tunings on the initial condition. 
	Constraint from cosmic strings is also relaxed.
\end{abstract}

\end{center}

\end{titlepage}

\newpage
%\tableofcontents

%\newpage
\vspace{1cm}

%%%%%%%%%%%%%%%%%%%%%%%%%%%%%%%%%%%%%%%%%%%%%%%%%%%%%%%%%%%%%%%%%%%%%%
\section{Introduction} \label{intro}
%%%%%%%%%%%%%%%%%%%%%%%%%%%%%%%%%%%%%%%%%%%%%%%%%%%%%%%%%%%%%%%%%%%%%%

The current cosmic microwave back ground (CMB) observations, such as the Wilkinson Microwave Anisotropy Probe (WMAP) observation~\cite{Hinshaw:2012fq}, 
strongly support the existence of an accelerated expansion era called inflation in the very early stage of the universe.
The inflation is driven by some scalar field, called inflaton, whose potential is nearly flat.
Unfortunately, such a scalar field does not exist in the framework of the well-established Standard Model of the particle physics, so we must go beyond the Standard Model.
One of the most plausible extensions of the Standard Model is the supersymmetry (SUSY) or supergravity.
Thus it is well-motivated to consider the inflation model in the framework of SUSY.

Up to now, many SUSY inflation models have been proposed~\cite{Yamaguchi:2011kg}.
Among these, the SUSY hybrid inflation is one of the simplest and most plausible models~\cite{Copeland:1994vg}.
In this model, the energy scale of the inflation, in terms of the Hubble scale during inflation, 
is required to be of the order of $10^{9}$--$10^{11}$\,GeV in order to reproduce the observed density perturbation.
The red-tilted power spectrum, $n_S \simeq 0.98$ ($n_S \simeq 0.96$ by relying on the non-minimal K\"ahler potential \cite{BasteroGil:2006cm}), can also be reproduced, which is supported by the CMB observation.

However, there is a drawback in SUSY hybrid inflation model.
In supergravity, we must include the constant term in the superpotential : $W_0 = m_{3/2}M_P^2$, where $m_{3/2}$ is the gravitino mass and $M_P$ is the reduced Planck scale, in order to cancel the SUSY breaking vacuum energy.
Including such a term, the linear term for the inflaton is induced in the potential
and this term may change the dynamics of the inflaton significantly~\cite{Buchmuller:2000zm,Senoguz:2004vu,Nakayama:2010xf}.
In particular, severe fine tuning on the initial condition is needed for successful inflation
and the allowed parameter region consistent with the WMAP observation shrinks for larger gravitino mass~\cite{Nakayama:2010xf}.
%This may be crucial if the model is embedded in some particle physics model like the SUSY axion model~\cite{Kawasaki:2010gv}.
In addition, the overproduction of thermally-produced gravitinos is also problematic because 
the reheating temperature tends to be high in the hybrid inflation model.
For these reasons, large gravitino mass of $m_{3/2}\gtrsim 100\,{\rm TeV}$ is disfavored in the SUSY hybrid inflation model.
On the other hand, recent observations of Higgs-boson-like particle at the LHC~\cite{:2012gk} may indicate a 
relatively high-scale SUSY~\cite{Okada:1990gg}. 
At first sight, therefore, the hybrid inflation model seems to be disfavored 
in the light of the recent LHC result.\footnote{%%
A variant model of the SUSY hybrid inflation, smooth hybrid inflation, 
allows large gravitino mass because the inflationary dynamics is less affected by the linear term as shown in \cite{Kawasaki:2012wj}.
}%%

In this paper we propose a modified model for hybrid inflation, where
the problematic constant term in the superpotential is replaced with a dynamical field.
It is dynamically set to a small value during hybrid inflation, 
which avoids the problem with linear term inflaton potential
in the original hybrid inflation model, and obtains a large vacuum expectation value (VEV) 
after inflation yielding a large gravitino mass.\footnote{%%
Another idea to avoid the linear term problem in hybrid inflation was proposed 
in Ref.~\cite{Higaki:2012iq} based on no-scale supergravity.
}
The added dynamical term in the superpotential is the same as that for new inflation 
model~\cite{Izawa:1996dv,Ibe:2006ck}.
A similar model was proposed in the context of preinflation for solving 
the severe initial value problem of new inflation~\cite{Izawa:1997df}, 
and also in the context of double inflation~\cite{Kawasaki:1997ju},
where a period of new inflation follows after hybrid inflation.
In the present purpose, we do not necessarily need a period of new inflation; 
it only guarantees the successful dynamics of hybrid inflation.
Actually we find that in such a setup, heavy gravitino scenario is suitably 
consistent with hybrid inflation model.

This paper is organized as follows.
In Sec.~\ref{model}, we introduce our inflation model and dynamics 
after inflation is considered.
In Sec.~\ref{constraint}, the model parameters are constrained from observations 
and the initial value problem for the inflaton and the gravitino problem are discussed.
Sec.~\ref{conc} is devoted to the conclusion.

%%%%%%%%%%%%%%%%%%%%%%%%%%%%%%%%%%%%%%%%%%%%%%%%%%%%%%%%%%%%%%%%%%%%%%
\section{Modified supersymmetric hybrid inflation model} \label{model}
%%%%%%%%%%%%%%%%%%%%%%%%%%%%%%%%%%%%%%%%%%%%%%%%%%%%%%%%%%%%%%%%%%%%%%

%%%%%%%%%%%%%%%%%%%%%%%%%%%%%%%%%%%%%%%%%%%%%%%%%%%%%%%%%%%%%%%%%%%%%%
\subsection{The inflaton potential} \label{potential}
%%%%%%%%%%%%%%%%%%%%%%%%%%%%%%%%%%%%%%%%%%%%%%%%%%%%%%%%%%%%%%%%%%%%%%

First we introduce a modified SUSY hybrid inflation model.
The superpotential in our model is given by
\be
	W = W_H + W_N.
	\label{superpot}
\ee
The first term is the superpotential for the hybrid inflation given by
\be
	W_H = \kappa S (\Psi\bar\Psi - M^2),
	\label{W_H}
\ee
where $S$ and $\Psi$ ($\bar\Psi$) are chiral superfields whose scalar components 
play roles of the inflaton and the waterfall field, 
$\kappa$ is a dimensionless coupling constant and $M$ gives the VEV of waterfall field.
We take the Planck unit, i.e. $M_P = 1$, throughout the paper. 
This model has $U(1)_R$ symmetry with charge assignments of $+2$, $0$ and $0$ 
for $S$, $\Psi$ and $\bar\Psi$ respectively. 
In addition, there is another $U(1)$ symmetry whose charge assignments are 
$0$, $+1$ and $-1$ for $S$, $\Psi$ and $\bar\Psi$ respectively. 
We assume that this is a gauge symmetry.\footnote{%%
It is possible that this U(1) is a global symmetry, such as Peccei-Quinn 
symmetry~\cite{Kawasaki:2010gv}.
}
The second term in (\ref{superpot}) is given by
\be
	W_N = \Phi \bigg( v^2 - \frac{g}{n+1} \Phi^n \bigg),
	\label{W_N}
\ee
where $\Phi$ is a chiral superfield, $v$ gives the energy scale of $\Phi$ potential,  
$g$ is a self coupling constant and $n$ is an integer larger than 2.
A discrete $R$-symmetry $Z^{R}_{2n}$ under which $\Phi$ has a charge $+2$ ensures 
this form of the superpotential.\footnote{%% 
It is also understood in the following way : $\Phi$ has a U(1)$_R$ charge $2/(n+1)$, 
while it couples to additional chiral matters which condensates at the dynamical 
SUSY breaking scale yielding $\Phi v^2$ term~\cite{Izawa:1997df}.
}
Note that this superpotential has the same form as the one for 
the new inflation~\cite{Izawa:1996dv}
and it becomes a non-zero constant term at the potential minimum, 
giving the gravitino mass.
The K\"ahler potential is taken to be
\be
	K = K_H + K_N
	\label{Kahler_pot}
\ee
where $K_H$ and $K_N$ are respectively given by
\be
	K_H = |S|^2 + |\Psi|^2 + |\bar\Psi|^2 + |\Phi|^2 
	+ \frac{k_S}{4} |S|^4 + k_1 |S|^2 |\Psi|^2 + k_2 |S|^2 |\bar\Psi|^2 
	+ \frac{k_{SS}}{6} |S|^6 + \dots,
	\label{K_H}
\ee
and
\be
	K_N = |\Phi|^2 + \frac{c_N}{4} |\Phi|^4 + \dots, 
	\label{K_N}
\ee
where $k_S$, $k_1$, $k_2$, $k_{SS}$ and $c_N$ are dimensionless coefficients and dots denote higher order Planck suppressed terms. In the following we assume $c_N$ is positive.

The F-term scalar potential is calculated from the formula
\be 
	V_F = e^{K} \big[ K^{ij^*} D_i W  D_{j^*} W^* - 3 |W|^2 \big],
\ee
where $D_i W = W_i + K_i W$ and the subscript represents derivative 
with respect to corresponding field and $K^{i j^*}=K_{i j^*}^{-1}$.
For $|S| > M$, hybrid inflation takes place and the waterfall fields are stabilized 
at the origin : $\Psi = \bar\Psi = 0$.
Thus we get the F-term scalar potential during inflation as 
\be
\begin{split}
	V = & \exp \bigg( |S|^2 + \frac{k_S}{4} |S|^4 + \frac{k_{SS}}{6} |S|^6 
	+ |\Phi|^2 + \frac{c_N}{4} |\Phi|^4 \bigg) \\[1mm]
	& \times \Bigg\{ 
	\bigg|  - \kappa M^2 \bigg( 1 + |S|^2 + \frac{k_S}{2}|S|^4 \bigg) 
	+ S^* \Phi \bigg( v^2 - \frac{g}{n+1} \Phi^n \bigg) \bigg|^2 \\[1mm]
	& + \frac{1}{1 + c_N |\Phi|^2} ~ 
	\bigg| v^2 \bigg( 1 + |\Phi|^2 + \frac{c_N}{2} |\Phi|^4 \bigg) 
	-g\Phi^n \bigg( 1 + \frac{|\Phi|^2}{n+1} + \frac{c_N |\Phi|^4}{2(n+1)} \bigg) \\[1mm]
	& - \kappa M^2 S \Phi^* \bigg( 1 + \frac{c_N}{2} |\Phi|^2 \bigg) \bigg|^2 
	- 3 \bigg| -\kappa M^2 S + \Phi \bigg( v^2 + \frac{g}{n+1} \Phi^n \bigg) \bigg|^2\Bigg\}.
	\end{split}
\ee
The scalar potential is conveniently divided into the following three pieces :
\be
	V = V_H + V_N + V_\mathrm{int}.
\ee
The each term is given by
\begin{align}
	&V_H = \kappa^2 M^4 \bigg( 1 -k_S |S|^2 + \frac{1}{2}\gamma|S|^4 \bigg) 
	+ V_\mathrm{CW}, \label{V_H} \\[2mm]
	&V_N = |v^2 + g\Phi^n|^2 - c_N v^4 |\Phi|^2, \label{V_N} \\[2mm]
	&V_\mathrm{int} = \kappa^2M^4|\Phi|^2+ \kappa M^2 v^2(S^*\Phi + \mathrm{c.c.}), \label{V_int}
\end{align}
where $\gamma = 1 - 7k_S/2 -3k_{SS} + 2k_S^2$ and we neglected the higher order 
Planck suppressed terms and assumed $v^2 \gg g \Phi^n$, and
$V_\mathrm{CW}$ is the Coleman-Weinberg effective potential~\cite{Coleman:1973jx} given by
\be
	V_\mathrm{CW} = \frac{\kappa^4 M^4}{32\pi^2} 
	\bigg[ (x^4 + 1) \ln \frac{x^4 -1 }{x^4} + 2x^2 \ln\frac{x^2 +1}{x^2-1} 
	+ 2 \ln\frac{\kappa^2 M^2 x^2}{\Lambda^2} -3 \bigg]
\ee
where $x \equiv |S| / M$ and $\Lambda$ is an ultraviolet cut-off scale.
In the limit of $|S| \gg M$, this is approximated as
\be
	V_\mathrm{CW} \simeq \frac{\kappa^4 M^4}{16 \pi^2} \ln \frac{\kappa^2 |S|^2}{\Lambda^2}.
	\label{CW_approx}
\ee

\begin{figure}[tp]
\centering
\includegraphics [width = 8cm, clip]{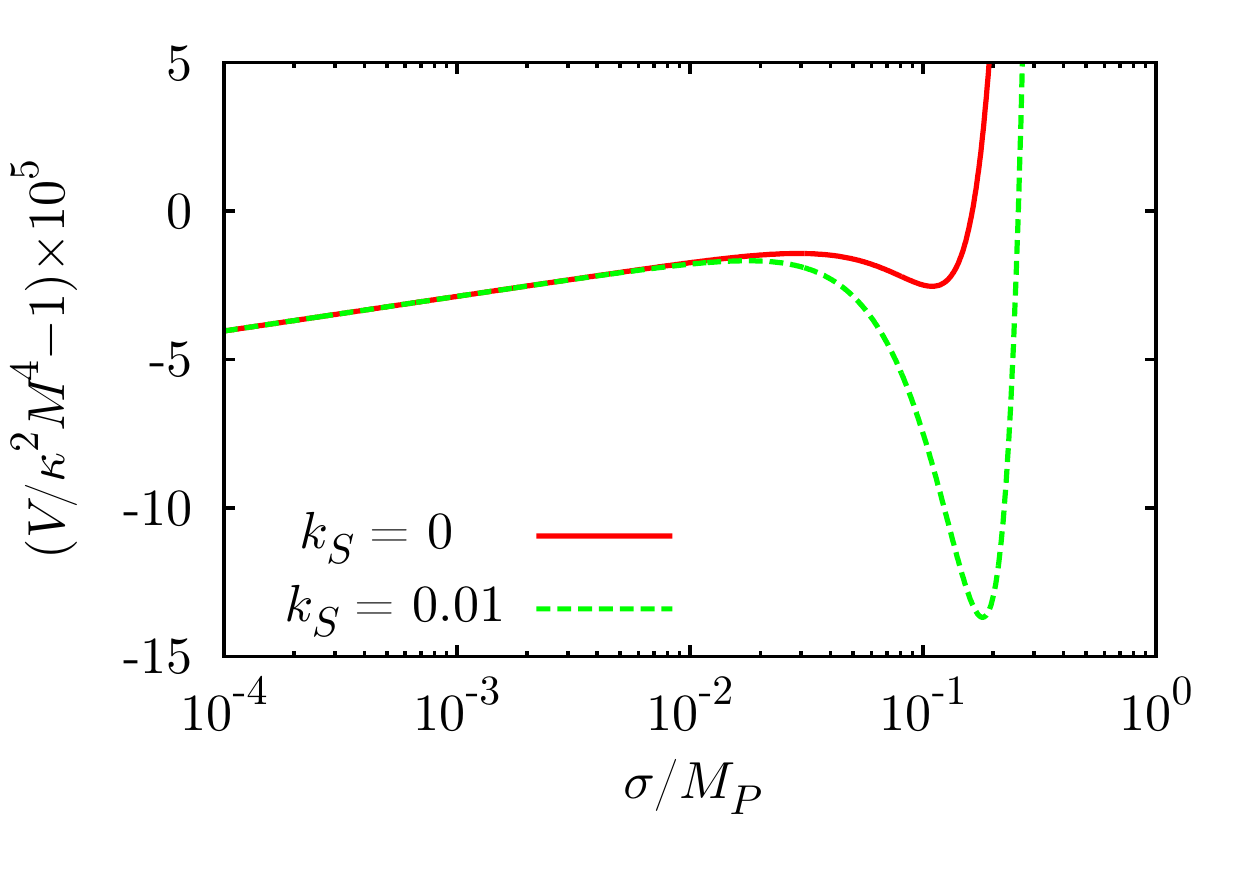}
\caption{
	The potential of the inflaton is shown.
	We have taken $M = 10^{15}~{\rm GeV}$, $\kappa = 0.02$, $v = 4 \times 10^{13}~{\rm GeV}$, $\Lambda = 10^{16}~{\rm GeV}$, 
	$k_S = 0$ (solid red) and $k_S = 0.01$ (dashed green).
}
\label{inflaton_pot}
\end{figure}

In the above calculation, we implicitly assumed that $\Phi$ is a subdominant component during inflation, i.e., $v \ll \sqrt{\kappa}M$.
%Because $V_\mathrm{int} \gg V_N$ is satisfied under this assumption, 
From (\ref{V_int}), the minimum of $\Phi$ during inflation is determined by
\be
	\Phi_\mathrm{min} = - \frac{v^2}{\kappa M^2} S,
\ee
which leads to the effective potential for $S$ : 
$V_\mathrm{int} (\Phi_\mathrm{min}) = -v^4 |S|^2$.
Then, defining the inflaton field as $\sigma = \sqrt{2}|S|$, the effective potential 
of the inflaton is derived as
\be
	V(\sigma) = \kappa^2 M^4 
	\bigg( 1 - \frac{1}{2}k_S \sigma^2 + \frac{1}{8}\gamma\sigma^4 \bigg)
	 - \frac{v^4}{2} \sigma^2 
	 + \frac{\kappa^4 M^4}{16 \pi^2} \ln \frac{\kappa^2 \sigma^2}{2\Lambda^2},
	\label{V_sigma}
\ee
where we used the approximation for the Coleman-Weinberg potential (\ref{CW_approx}).
Note that there is no dangerous linear term in the potential which may spoil 
the inflaton dynamics.
The inflaton potential (\ref{V_sigma}) is illustrated in Fig.~\ref{inflaton_pot}.
The slow-roll parameters are calculated as
\begin{align}
	\epsilon &\equiv \frac{1}{2}\bigg( \frac{V'}{V} \bigg)^2 
	= \frac{1}{2}\bigg[-\bigg(k_S + \frac{v^4}{\kappa^2 M^4}\bigg) \sigma 
	+ \frac{1}{2}\gamma\sigma^3 + \frac{\kappa^2}{8\pi^2\sigma}\bigg]^2 \\[2mm]
	\eta &\equiv \frac{V''}{V} 
	= -k_S - \frac{v^4}{\kappa^2 M^4} + \frac{3}{2} \gamma\sigma^2 
	-\frac{\kappa^2}{8\pi^2\sigma^2},
\end{align}
where the prime denotes the derivative with respect to the inflaton field.
The inflation lasts as long as the slow-roll conditions, $\epsilon \ll 1$ 
and $|\eta| \ll 1$, are satisfied 
and ends when the inflaton reaches the value where the slow-roll condition breaks down or 
the waterfall point given by $\sigma_{\rm wf} = \sqrt{2}M$ where the waterfall fields 
become tachyonic.
Thus the value of $\sigma$ at the end of the hybrid inflation is given by
\be
	\sigma_f = \max\{ \sigma_c, \sqrt{2}M \}, 
	~~~\text{where}~~~\sigma_c \simeq \frac{\kappa}{2\sqrt{2}\pi}.
\ee
%%
%We will constrain our model parameters by imposing the observational values after we consider the dynamics of $\Phi$ after inflation and derive the relation of $v$ with $m_{3/2}$ in next subsection.

%Imposing the WMAP normalization, the constraint on the paramter space is shown as the small-dotted magenta lines in Fig~\ref{Fig1}.
%This is also shown as the small-dotted magenta lines in Fig~\ref{Fig2}.

%%%%%%%%%%%%%%%%%%%%%%%%%%% FIGURE  %%%%%%%%%%%%%%%%%%%%%%%%%%%%%%%%%%%%%%
%%%%%%%%%%%%%%%%%%%%%%%%%%%%%%%%%%%%%%%%%%%%%%%%%%%%%%%%%%%%%%%%%%%%%%

%%%%%%%%%%%%%%%%%%%%%%%%%%%%%%%%%%%%%%%%%%%%%%%%%%%%%%%%%%%%%%%%%%%%%%
\subsection{Dynamics after inflation} \label{dynamics}
%%%%%%%%%%%%%%%%%%%%%%%%%%%%%%%%%%%%%%%%%%%%%%%%%%%%%%%%%%%%%%%%%%%%%%

\begin{figure}[tp]
\centering
  \subfigure{\includegraphics[width = 7.5cm, clip]{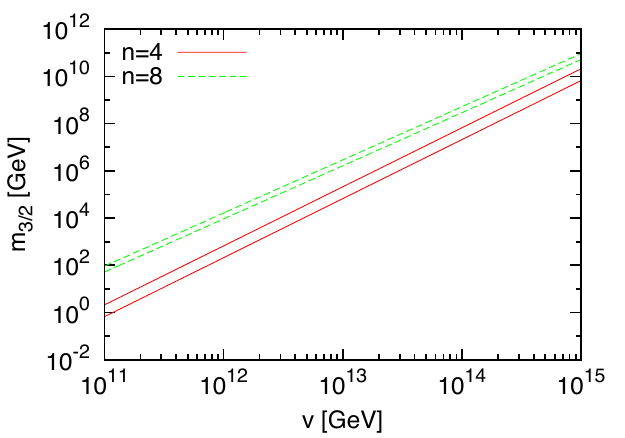}}
  \subfigure{\includegraphics[width = 7.5cm, clip]{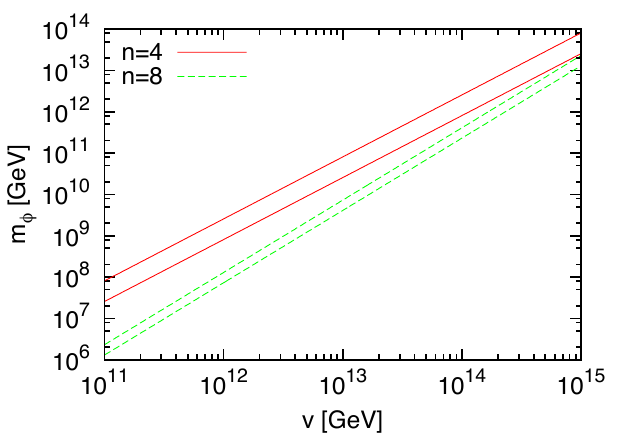}}
  \label{mgrav}
  \caption{
	Gravitino mass (left) and $\Phi$ mass (right) as a function of $v$ 
	for $n=4$ (red, solid) and $n=8$ (green, dashed).
	Two lines correspond to $g=10^{-2}$ (upper) and $g=1$ (lower) in the left panel 
	and $g=10^{-2}$ (lower) and $g=1$ (upper) in the right panel.
	}
\label{fig:mgrav}
\end{figure}

After the end of inflation, the inflaton and waterfall fields starts to oscillate, 
so the universe is dominated by oscillating scalar fields which behave like matter.
On the other hand, $\Phi$ also starts to oscillate around the origin 
due to the Hubble-induced mass.
%Defining $\phi = \sqrt{2} {\rm Re}(\Phi)$, 
The initial amplitude of $\Phi$-oscillation is given by the value of $\Phi$ 
at the end of inflation : 
\be
	\Phi_f \equiv -\frac{v^2 \sigma_f}{\sqrt{2}\kappa M^2}.
\ee
Because the total energy density is dominated by the oscillating scalar fields, 
the Hubble-induced mass for $\Phi$ is given by $m^2_\phi = (3/2)H^2$.
So the amplitude of $\Phi$-oscillation decreases proportional to $a^{-3/4}$ 
($a$ is the scale factor of the cosmic expansion) in this era.
Since $V_N \sim v^4$ for $\Phi \sim 0$ while $V_H \propto a^{-3}$, 
$V_N$ starts to dominate the whole potential at $H \sim v^2/\sqrt{3}$.
Since the mass-squared of $\Phi$ around the origin is given by $\sim -\sqrt{c_N} v^2$,
it starts to roll down and oscillate around the true minimum $\Phi_{0}$ 
as soon as it comes to dominate the Universe,
where
\be
	|\Phi_{0}|^n = \frac{v^2}{g},
\ee
which corresponds to $D_\Phi W = 0$.\footnote{%%
Since $|\Phi_{\rm min}|$ is much larger than $H_{\rm inf} \equiv \kappa M^2/\sqrt{3}$, 
which is typical scale of the field fluctuation, $\Phi$ can be regarded 
as a homogeneous field. 
Thus $\Phi$ rolls down to one of the minima of the potential in whole region 
of the observable Universe at $H\sim v^2/\sqrt{3}$ 
and no domain walls are formed. 
}
Although the second inflation could take place if $c_N \ll1$, we do not require it.
At the minimum, $V_N$ becomes negative and its vacuum energy is given by
\be
	V_{\rm vac} = -3 e^K |W|^2 \simeq -3 \bigg( \frac{n}{n+1} \bigg)^2 v^4 |\Phi_{0}|^2.
\ee
Requiring that $V_{\rm vac}$ compensates the vacuum energy from the SUSY breaking sector, denoted by $\Lambda^4_{\rm SUSY}$, we get the gravitino mass as follows : 
\be
	m_{3/2} = \frac{\Lambda^2_{\rm SUSY}}{\sqrt{3}} 
	= \bigg( \frac{n}{n+1} \bigg) v^2 \bigg(\frac{v^2}{g} \bigg)^{1/n}.
\ee
The mass of $\Phi$ around the minimum is given by
\be
	m_\phi = ng^{1/n}v^{2-2/n} \simeq ng^{\frac{2}{n+1}} m_{3/2}^{\frac{n-1}{n+1}}.
\ee
Fig.~\ref{fig:mgrav} shows gravitino mass and $\Phi$ mass as a function of $v$ for $n=4$ (red, solid) and $n=8$ (green, dashed).
Two lines correspond to $g=10^{-2}$ (upper) and $g=1$ (lower) in the left panel and
$g=10^{-2}$ (lower) and $g=1$ (upper) in the right panel.
From this figure, it is seen that we should have $v\sim 10^{12}$--$10^{13}$\,GeV for $m_{3/2}=100$--$1000$\,TeV.

%%%%%%%%%%%%%%%%%%%%%%% FIGURE  %%%%%%%%%%%%%%%%%%%%%%%%%%%%%%%%%%%%%%
%%%%%%%%%%%%%%%%%%%%%%%%%%%%%%%%%%%%%%%%%%%%%%%%%%%%%%%%%%%%%%%%%%

%%%%%%%%%%%%%%%%%%%%%%%%%%%%%%%%%%%%%%%%%%%%%%%%%%%%%%%%%%%%%%%%%%%%%%
\section{Constraints on the modified hybrid inflation model} \label{constraint}
%%%%%%%%%%%%%%%%%%%%%%%%%%%%%%%%%%%%%%%%%%%%%%%%%%%%%%%%%%%%%%%%%%%%%%

%%%%%%%%%%%%%%%%%%%%%%%%%%%%%%%%%%%%%%%%%%%%%%%%%%%%%%%%%%%%%%%%%%%%%%
\subsection{WMAP normalization and spectral index} \label{WMAP}
%%%%%%%%%%%%%%%%%%%%%%%%%%%%%%%%%%%%%%%%%%%%%%%%%%%%%%%%%%%%%%%%%%%%%%

\begin{figure}[tp]
\centering
\subfigure[$n = 4$, $g=1$ and $k_S = 0$]{
\includegraphics [width = 7.5cm, clip]{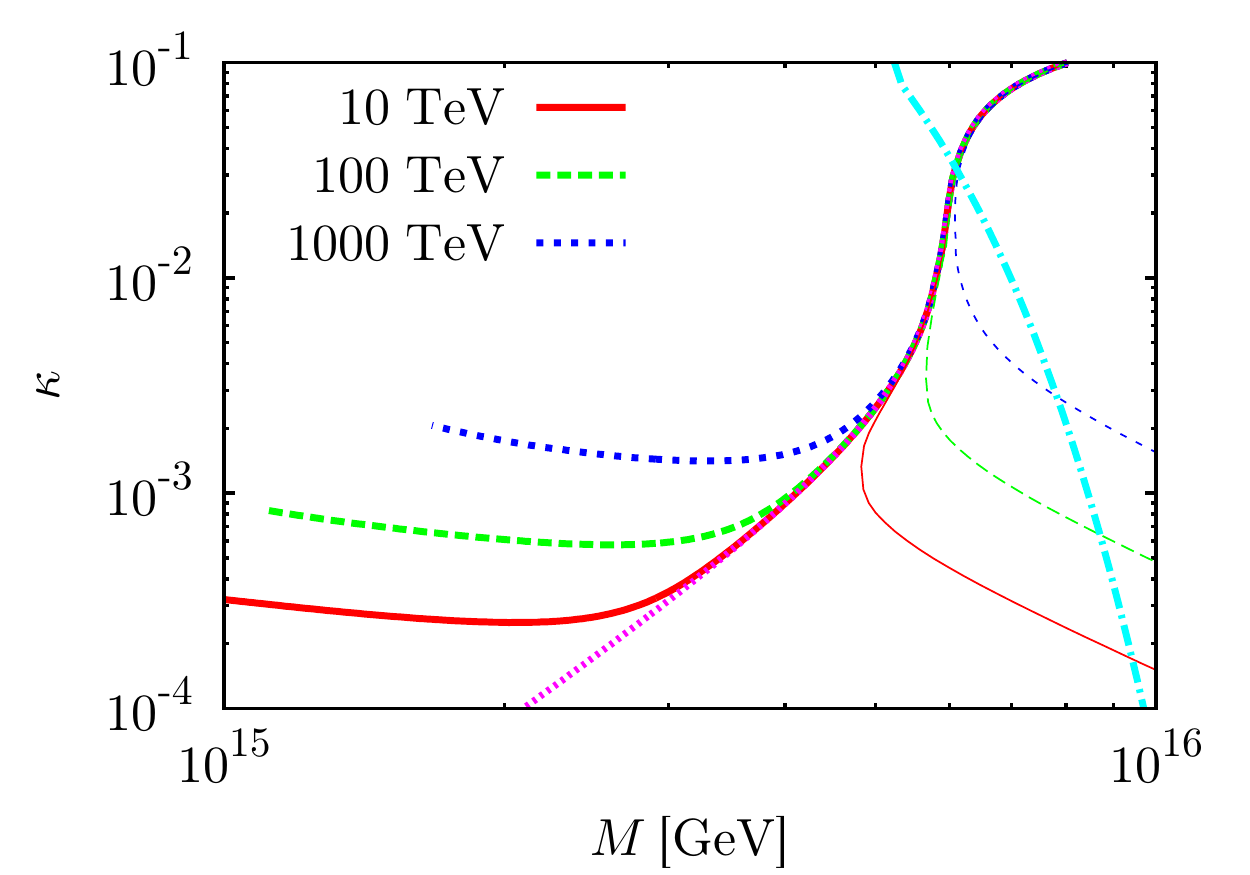}
\label{Fig1a}
}
\subfigure[$n = 4$, $g=0.01$ and $k_S = 0$]{
\includegraphics [width = 7.5cm, clip]{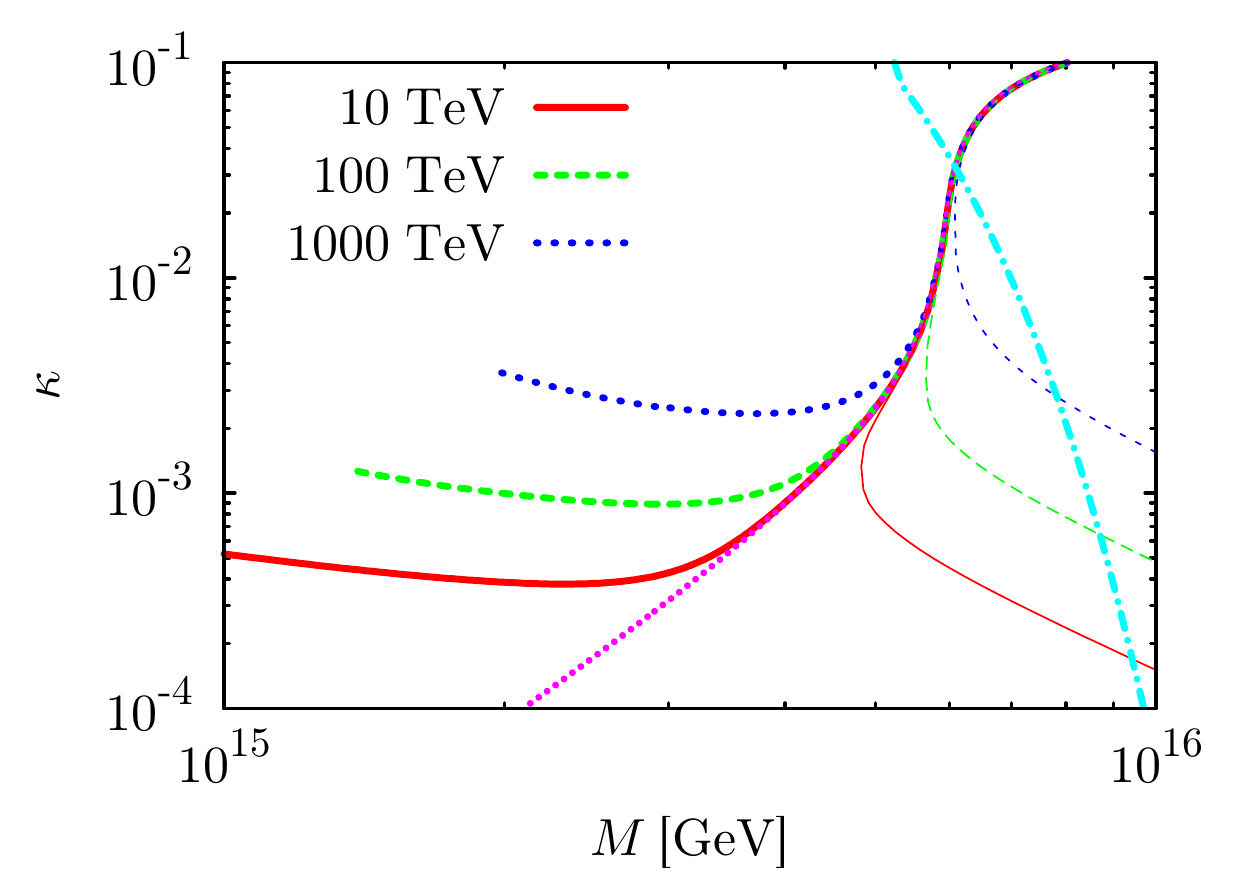}
\label{Fig1b}
}
\subfigure[$n = 6$, $g=1$ and $k_S = 0$]{
\includegraphics [width = 7.5cm, clip]{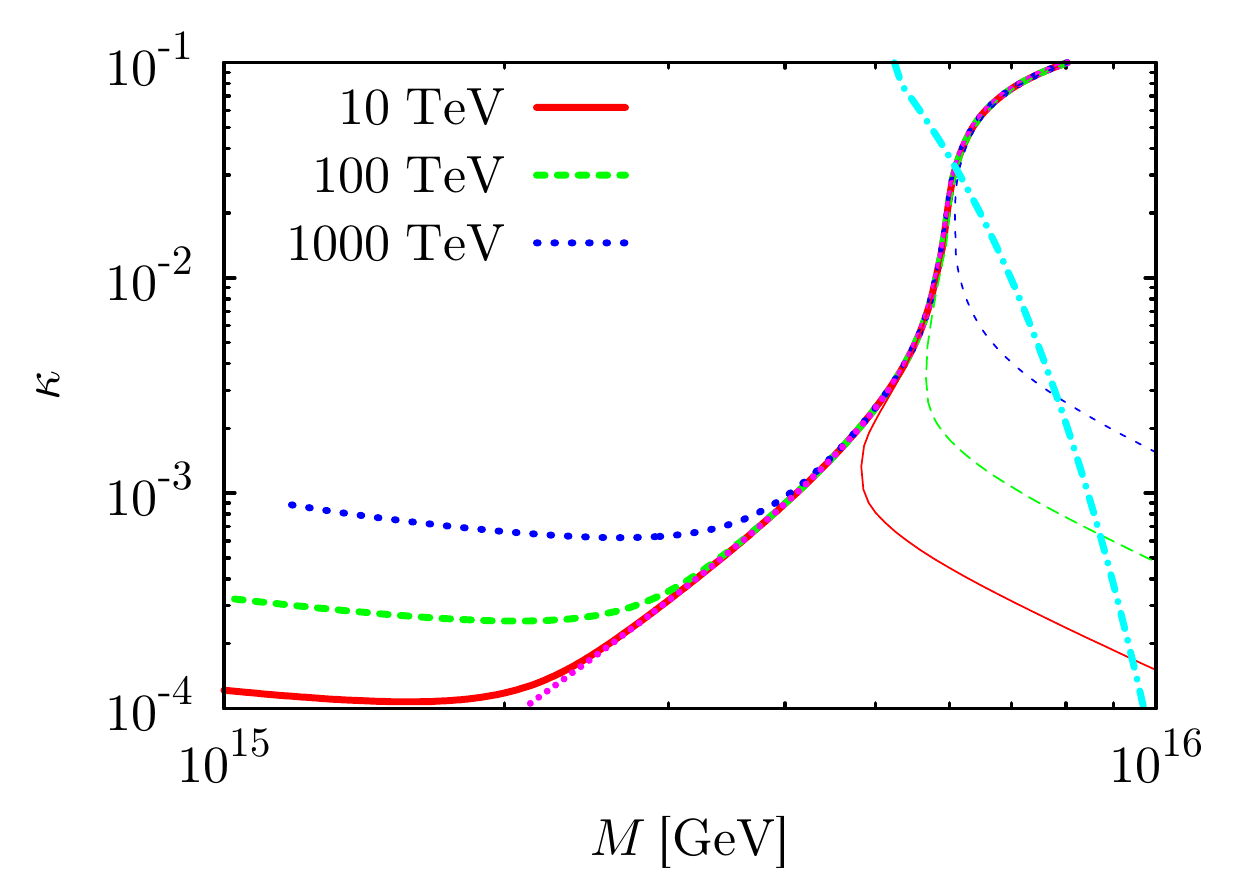}
\label{Fig1c}
}
\subfigure[$n = 4$, $g=1$ and $k_S = 0.01$]{
\includegraphics [width = 7.5cm, clip]{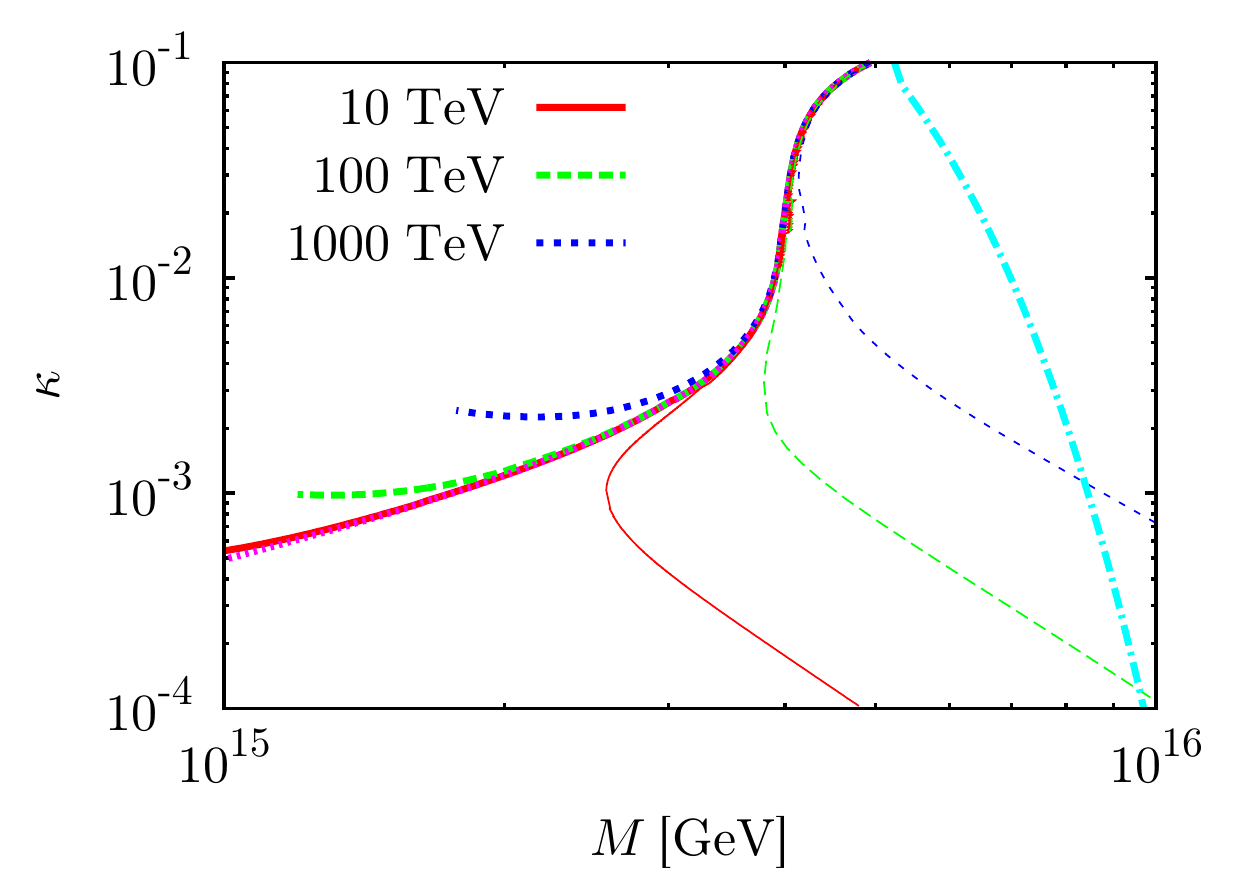}
\label{Fig1d}
}
\caption{
The allowed parameters satisfying the WMAP normalization are shown.
They are on the thick contours.
We have taken $n=4$ (Fig.~\ref{Fig1a}, \ref{Fig1c}, \ref{Fig1d}), 
$n=6$ (Fig.~\ref{Fig1c}), $k_S = 0$ (Fig.~\ref{Fig1a}, \ref{Fig1b}, \ref{Fig1c}), 
$k_S = 0.01$ (Fig.~\ref{Fig1d}) 
and $m_{3/2} = 10\,{\rm TeV}$ (solid red lines), $m_{3/2} = 100\,{\rm TeV}$ 
(dashed green lines) and $m_{3/2} = 1000\,{\rm TeV}$ (dotted blue lines).
Small-dotted magenta lines correspond to $m_{3/2} = 0$ and dashed-and-dotted cyan 
lines correspond to the upper bound on $\kappa$ from the cosmic string constraint.
Thin contours correspond to a ``traditional'' model including a constant term 
in the superpotential.
}
\label{Fig1}
\end{figure}

\begin{figure}[tp]
\centering
\subfigure[$n = 4$, $g=1$ and $k_S = 0$]{
\includegraphics [width = 7.5cm, clip]{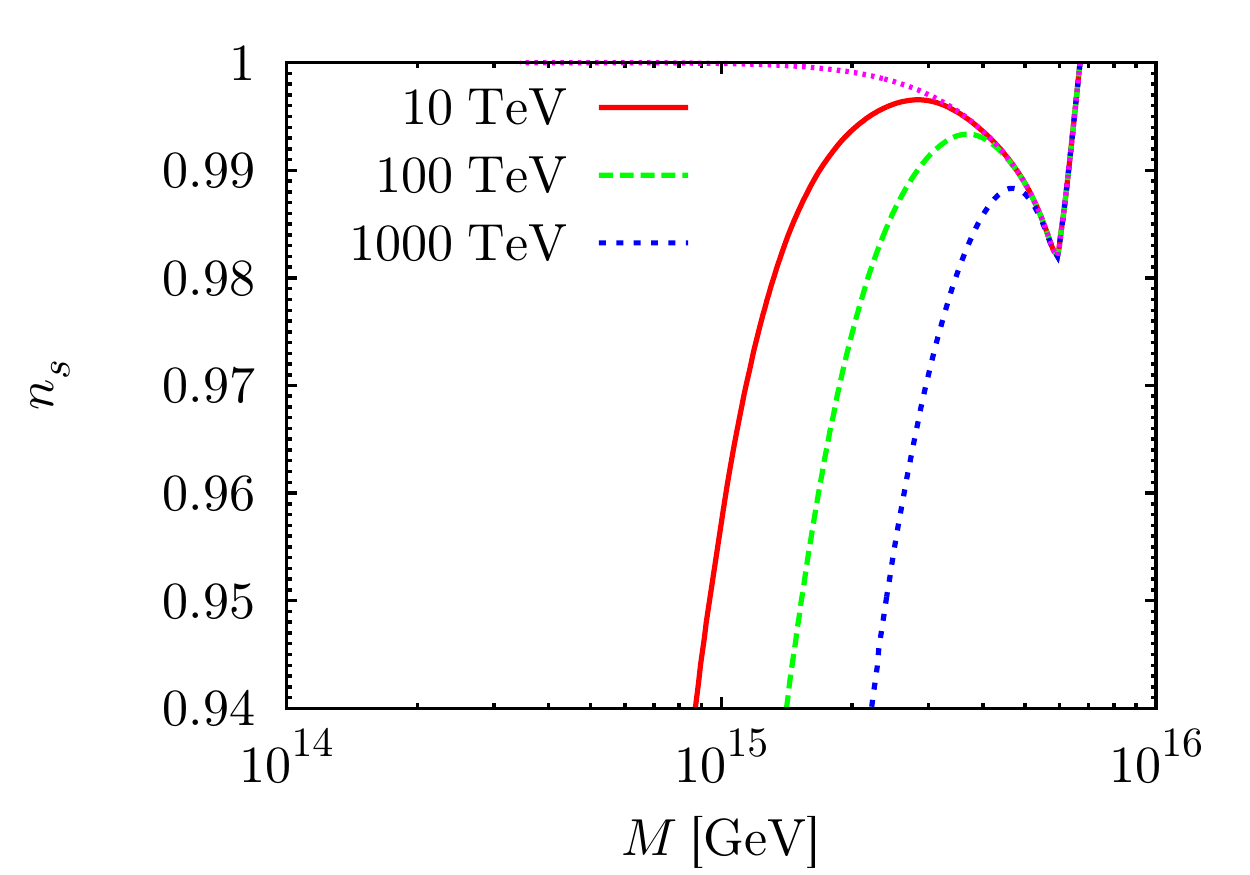}
\label{Fig2a}
}
\subfigure[$n = 4$, $g=0.01$ and $k_S = 0$]{
\includegraphics [width = 7.5cm, clip]{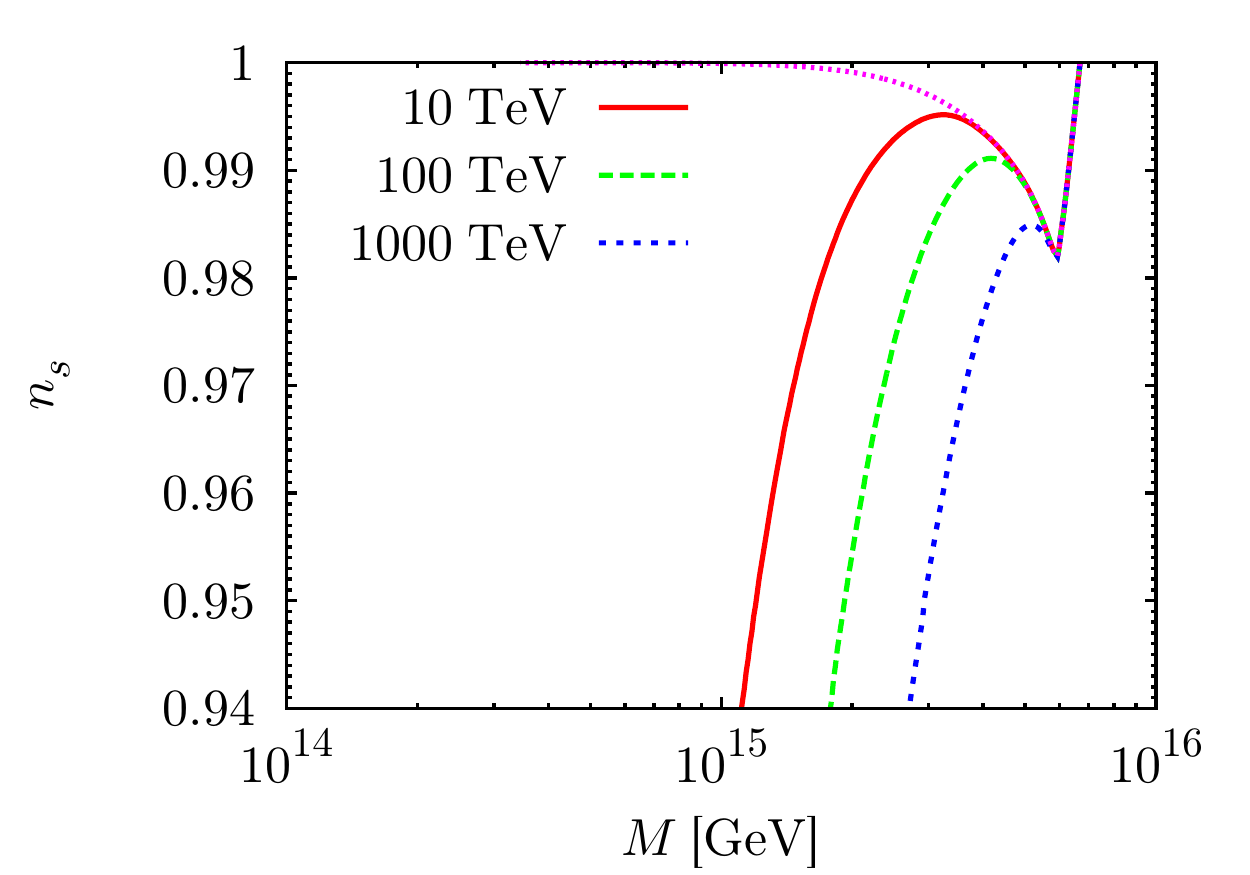}
\label{Fig2b}
}
\subfigure[$n = 6$, $g=1$ and $k_S = 0$]{
\includegraphics [width = 7.5cm, clip]{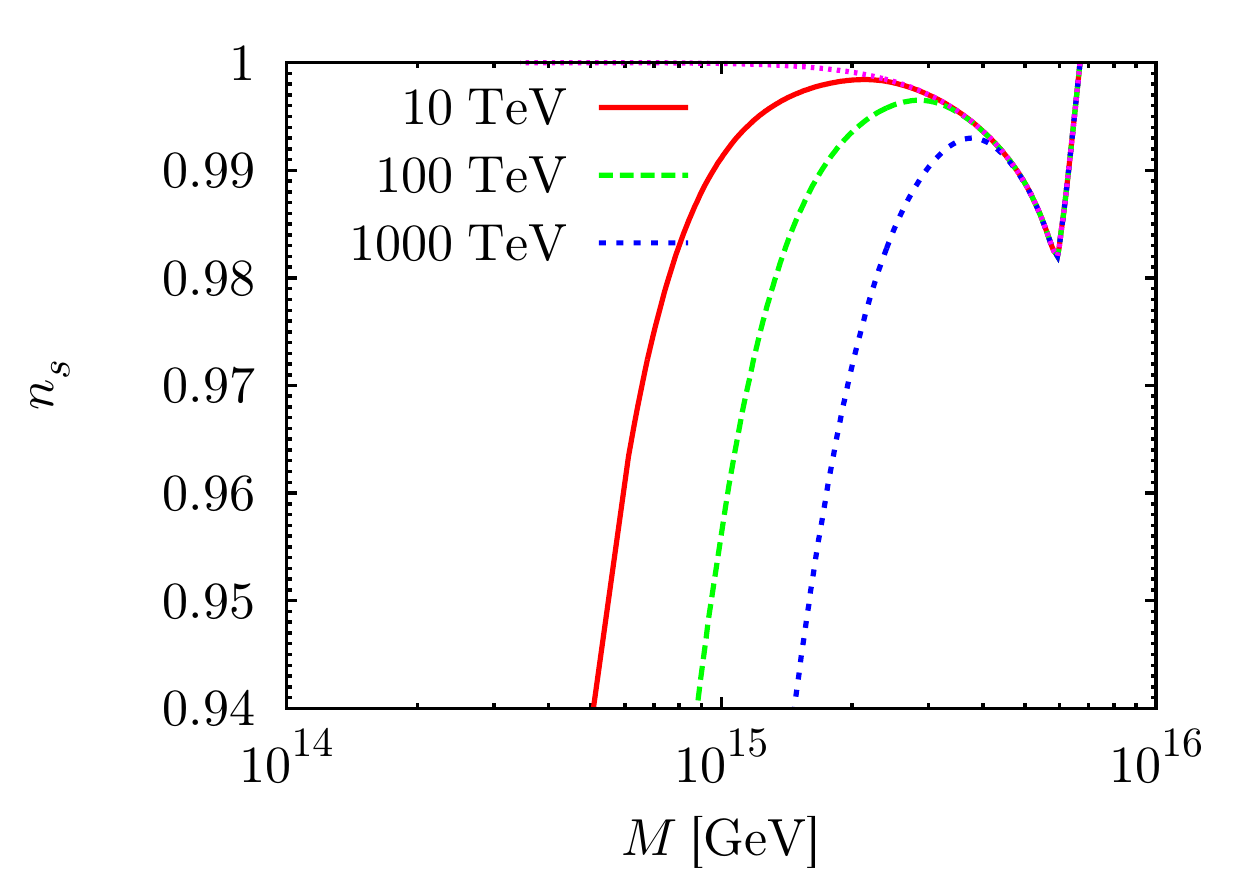}
\label{Fig2c}
}
\subfigure[$n = 4$, $g=1$ and $k_S = 0.01$]{
\includegraphics [width = 7.5cm, clip]{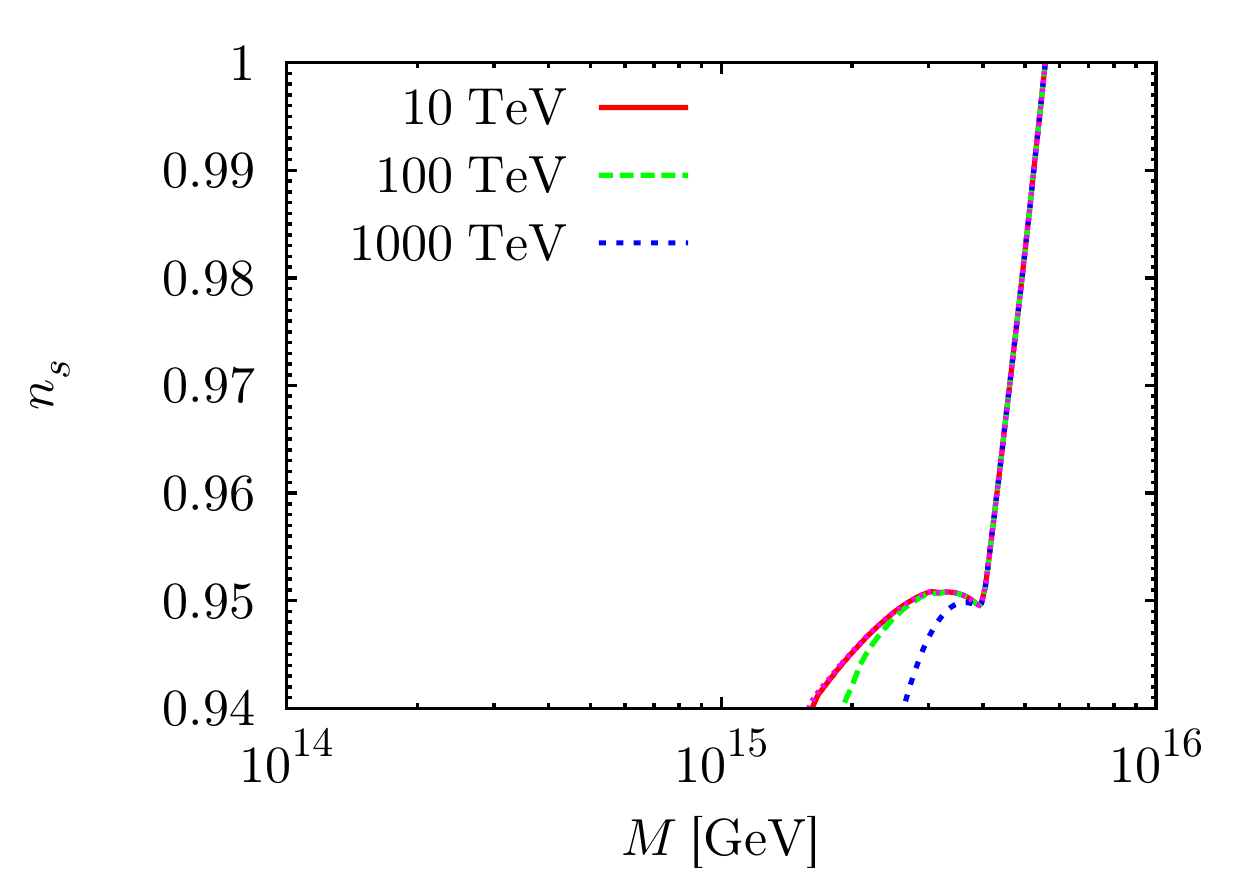}
\label{Fig2d}
}

\caption{
The spectral index $n_s$ as a function of $M$ imposing the WMAP normalization.
We have taken $n=4$ (Fig.~\ref{Fig2a}, \ref{Fig2b}, \ref{Fig2d}), 
$n=6$ (Fig.~\ref{Fig2c}), $k_S = 0$ (Fig.~\ref{Fig2a}, \ref{Fig2b}, \ref{Fig2c}), 
$k_S = 0.01$ (Fig.~\ref{Fig2d}) 
and $m_{3/2} = 10~{\rm TeV}$ (solid red lines), $m_{3/2} = 100~{\rm TeV}$ 
(dashed green lines) and $m_{3/2} = 1000~{\rm TeV}$ (dotted blue lines).
The small-dotted magenta lines correspond to $m_{3/2} = 0$.
}
\label{Fig2}
\end{figure}

Now let us constrain our model parameters from the CMB observation.
The WMAP observation gives the normalization to the power spectrum of 
the curvature perturbation 
as $\mathcal{P_R} \simeq 2.4 \times 10^{-9}$~\cite{Hinshaw:2012fq}.
The scalar spectral index $n_s$ is also constrained as $n_s = 0.968 \pm 0.012$.
These quantities are calculated as
\be
	\mathcal{P_R} = \frac{V}{24 \pi^2 \epsilon}
\ee
and
\be
	n_s \simeq 1-6\epsilon + 2\eta,
\ee
where $\epsilon$ and $\eta$ are the ones evaluated when the pivot scale $k_p = 0.002\,\mathrm{Mpc}^{-1}$ exits the horizon~\cite{Liddle&Lyth}.
The allowed parameters satisfying the WMAP normalization lies on the thick contours 
in Fig.~\ref{Fig1}. 
We have taken $n=4$, $g=1$ and $k_S=0$ in Fig.~\ref{Fig1a}, $n=4$, $g=0.01$ 
and $k_S=0$ in Fig.~\ref{Fig1b}, $n=6$, $g=1$ and $k_S=0$ in Fig.~\ref{Fig1d}, and 
$n=4$, $g=1$ and $k_S=0.01$ in Fig.~\ref{Fig1d}.
In each panel, we have taken $m_{3/2} = 10\,{\rm TeV}$ (solid red line), 
$100\,{\rm TeV}$ (dashed green line), $1000\,{\rm TeV}$ (dotted blue line) 
and $m_{3/2}=0$ (small-dotted magenta line).
For comparison, we show the case of ``traditional'' model including the constant term 
in the superpotential~\cite{Nakayama:2010xf} by thin contours.
We can see that $M$ can be smaller value down to $M \sim 2 \times 10^{15}\,{\rm GeV}$ 
even for $m_{3/2} \gtrsim 100\,{\rm TeV}$, which is contrasted to the traditional model. 
Fig.~\ref{Fig2} shows $n_s$ as a function of $M$ with the WMAP normalization is imposed.
Parameters are same as the ones used in Fig.~\ref{Fig1}.
We found that $n_s$ can be reduced to the WMAP central value without relying 
on the non-minimal K\"ahler potential.
This is due to the existence of the negative quadratic term 
in the potential~(\ref{V_sigma}) which comes from $V_{\rm int}$.
For smaller $M$, the condition $v\ll \sqrt{\kappa} M$ becomes violated and 
the dynamics of hybrid inflation is spoiled by this term.

%%%%%%%%%%%%%%%%%%%%%%% FIGURE  %%%%%%%%%%%%%%%%%%%%%%%%%%%%%%%%%%%%%%
%%%%%%%%%%%%%%%%%%%%%%%%%%%%%%%%%%%%%%%%%%%%%%%%%%%%%%%%%%%%%%%%%%

%%%%%%%%%%%%%%%%%%%%%%% FIGURE  %%%%%%%%%%%%%%%%%%%%%%%%%%%%%%%%%%%%%%
%%%%%%%%%%%%%%%%%%%%%%%%%%%%%%%%%%%%%%%%%%%%%%%%%%%%%%%%%%%%%%%%%%%%%%

%%%%%%%%%%%%%%%%%%%%%%%%%%%%%%%%%%%%%%%%%%%%%%%%%%%%%%%%%%%%%%%%%%%%%%
\subsection{Constraint from cosmic string} \label{cosmic_string}
%%%%%%%%%%%%%%%%%%%%%%%%%%%%%%%%%%%%%%%%%%%%%%%%%%%%%%%%%%%%%%%%%%%%%%

Soon after the end of hybrid inflation, the waterfall fields get large vacuum 
expectation values which spontaneously breaks $U(1)$ symmetry 
and hence the cosmic strings are inevitably formed~\cite{Jeannerot:2003qv}.
The tension of the cosmic string, denoted as $\mu$, is given 
by~\cite{Hill:1987ye,Jeannerot:2005mc}
\be
	\mu = 2 \pi M^2 \epsilon(\beta)
\ee
where $\beta = \kappa^2/2g^2$, $g^2 = 4\pi/25$ in grand unification models 
and $\epsilon(\beta)$ is defined through
\be
	\epsilon (\beta) = \begin{cases}
		1.04~\beta^{0.195} ~~~&\text{for}~~~\beta > 10^{-2} \\[1mm]
		\cfrac{2.4}{\ln (2/\beta)} ~~~&\text{for}~~~\beta < 10^{-2}.
	\end{cases}
\ee
Cosmic strings can be a source of large scale structure in the Universe 
in addition to the primordial density perturbation from the inflation.
The CMB observations constrain the tension of the cosmic string as 
$G\mu < (2-7) \times 10^{-7}$~\cite{Battye:2010xz}.
This gives a constraint on the parameter region as shown in Fig.~\ref{Fig1}, 
where the allowed parameter region is below the dashed-and-dotted cyan lines.

%%%%%%%%%%%%%%%%%%%%%%%%%%%%%%%%%%%%%%%%%%%%%%%%%%%%%%%%%%%%%%%%%%%%%%
\subsection{Initial condition} \label{initial condition}
%%%%%%%%%%%%%%%%%%%%%%%%%%%%%%%%%%%%%%%%%%%%%%%%%%%%%%%%%%%%%%%%%%%%%%

Let us consider the initial condition for the inflaton.\footnote{%%
Here, we assume that the initial value of the inflaton is placed on the inflationary trajectory since our interest is focused on the dynamics on it.
This assumption is reasonable because of the attractor behavior of the inflationary trajectory as shown in \cite{Clesse:2009ur}.
}%%
It is known that the initial condition for inflaton is constrained 
in the SUSY hybrid inflation model due to the constant term in the superpotential or 
non-minimal K\"ahler potential to reproduce the observed spectral 
index~\cite{Nakayama:2010xf}.

\begin{figure}[tp]
\centering
\subfigure[$m_{3/2} = 100~{\rm TeV}$, $g=1$ and $k_S = 0$]{
	\includegraphics [width = 7.5cm, clip]{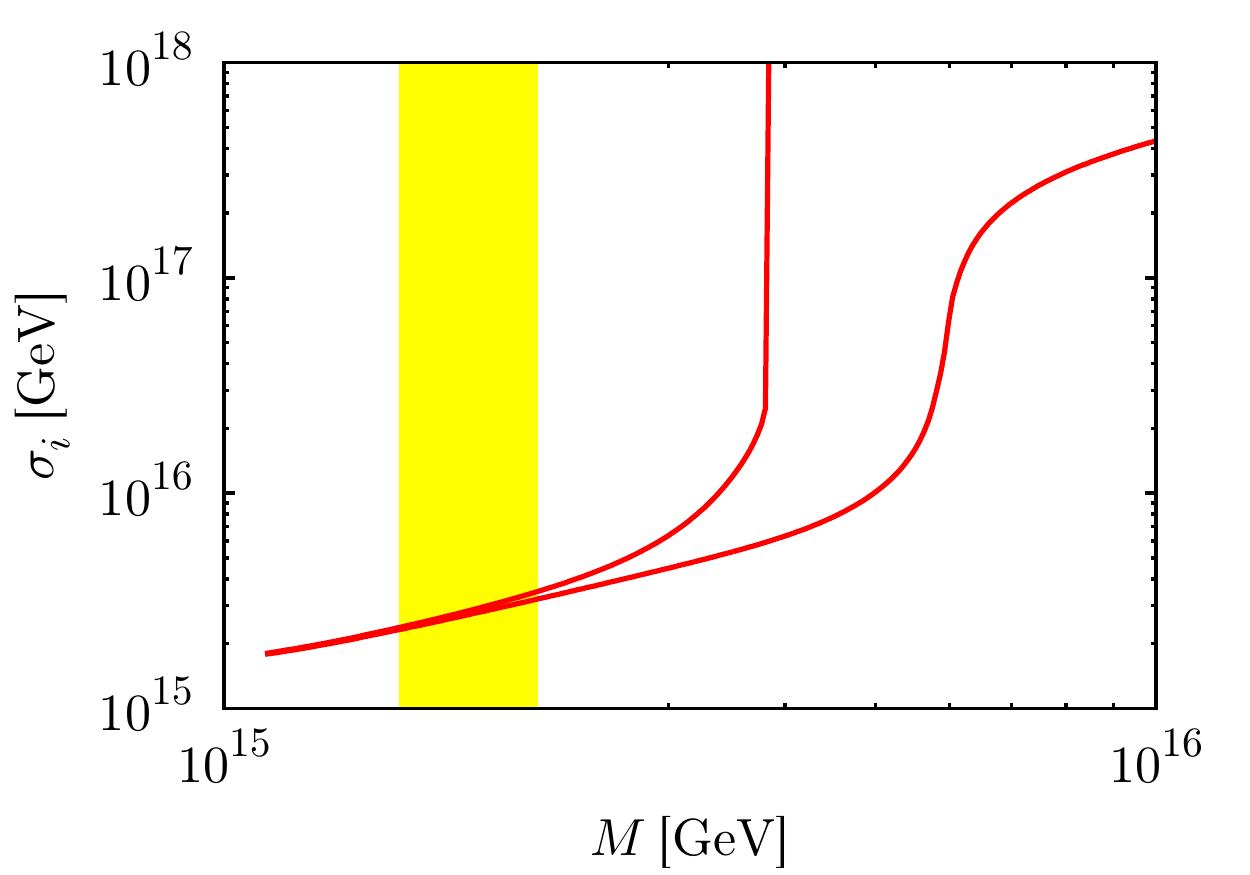}
	\label{Fig4a}
}
\subfigure[$m_{3/2} = 1000~{\rm TeV}$, $g=1$ and $k_S = 0$]{
	\includegraphics [width = 7.5cm, clip]{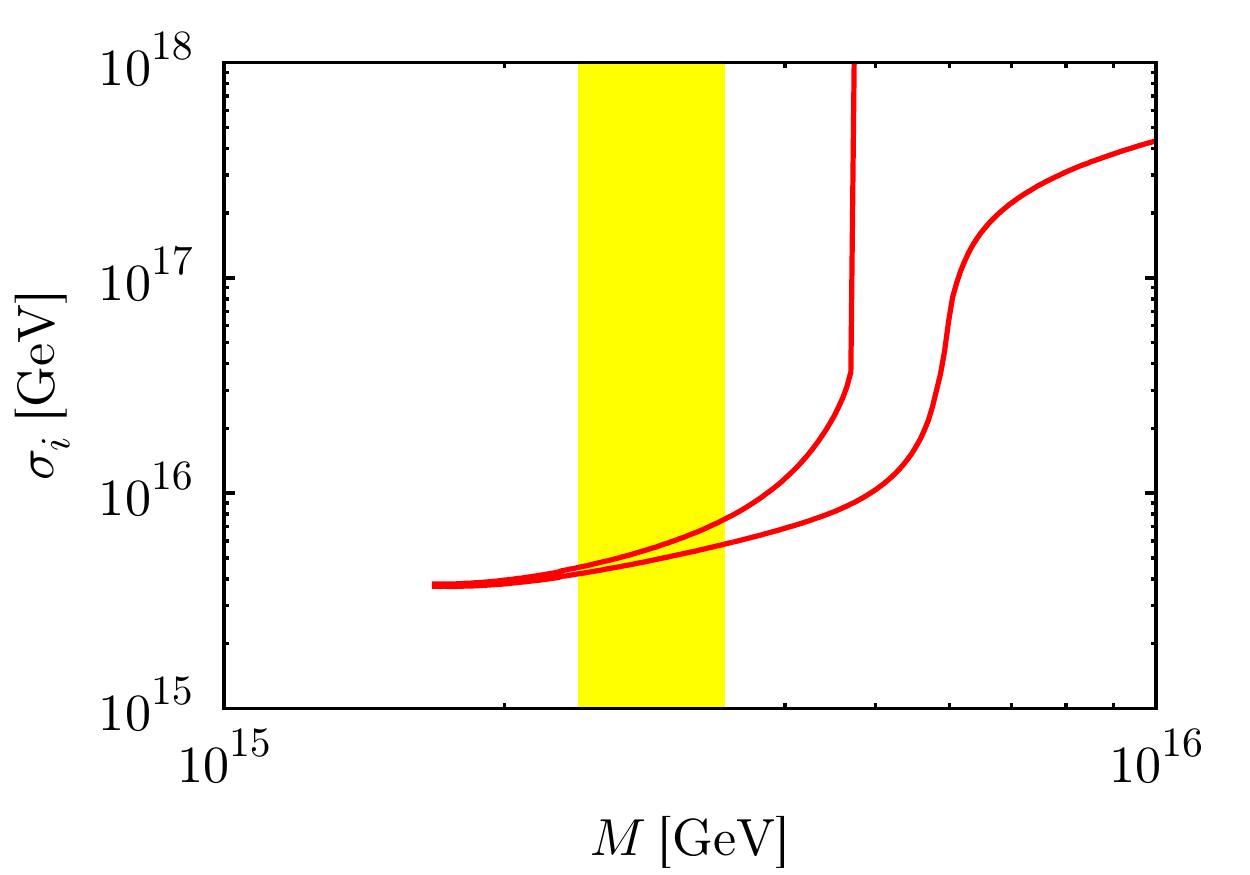}
	\label{Fig4b}
}
\subfigure[$m_{3/2} = 100~{\rm TeV}$, $g=0.01$ and $k_S = 0$]{
	\includegraphics [width = 7.5cm, clip]{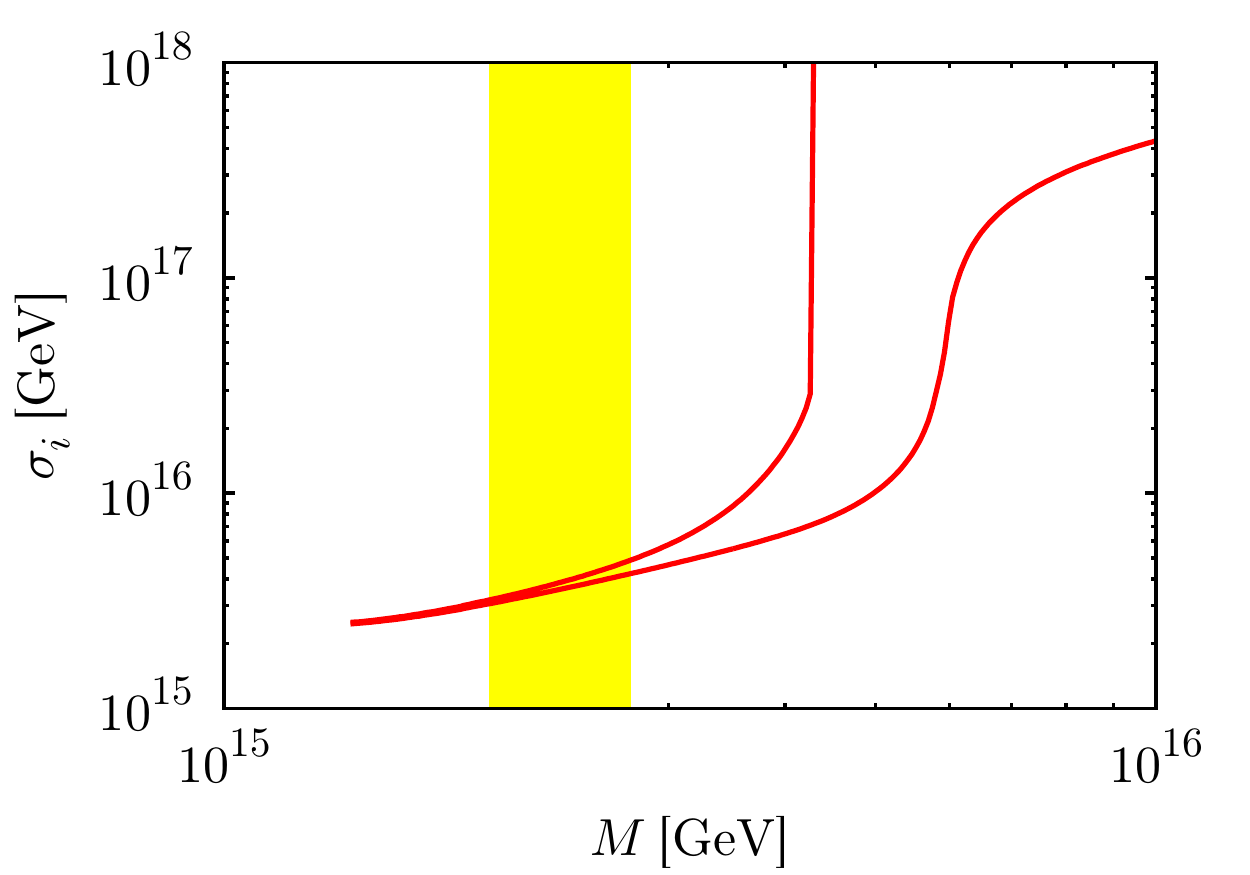}
	\label{Fig4c}
}
\subfigure[$m_{3/2} = 1000~{\rm TeV}$, $g=0.01$ and $k_S = 0$]{
	\includegraphics [width = 7.5cm, clip]{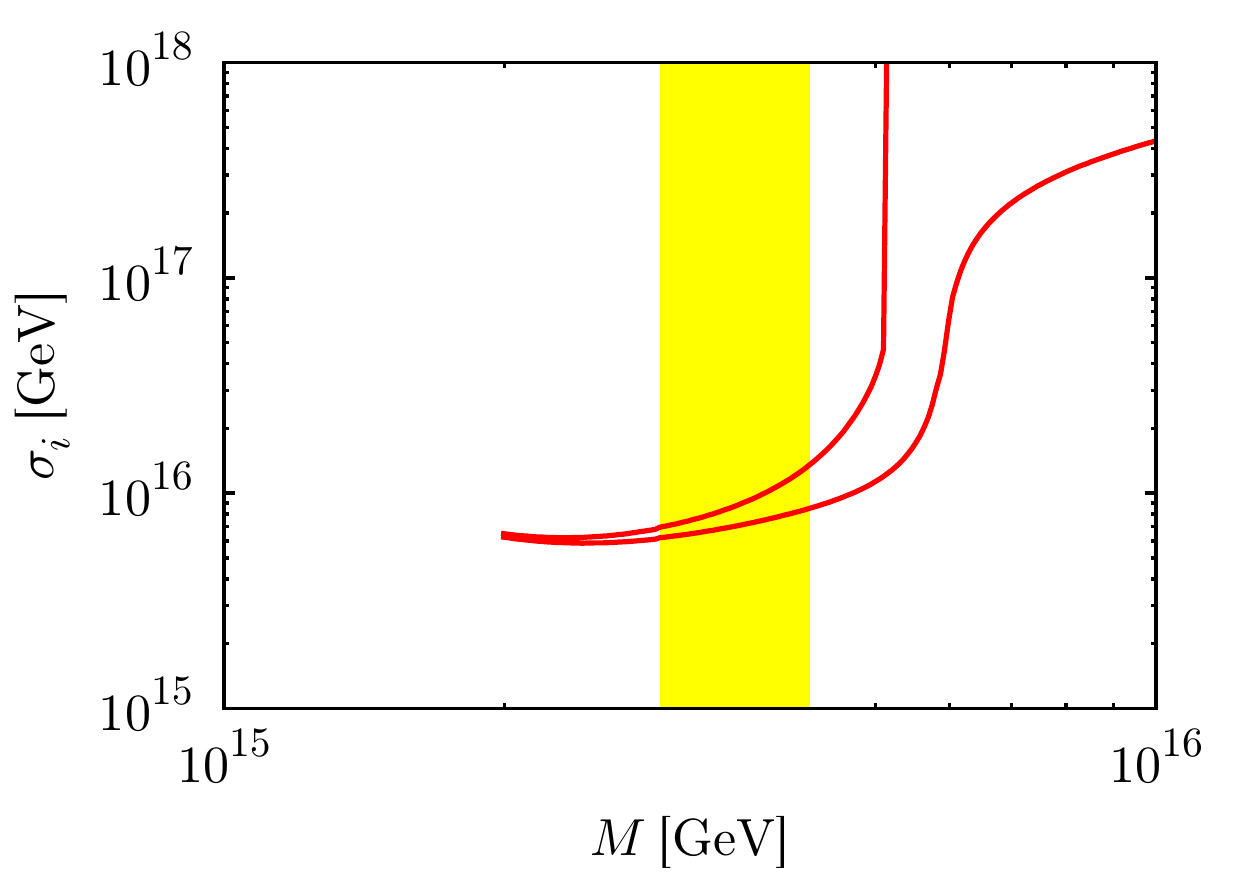}
	\label{Fig4d}
}
\subfigure[$m_{3/2} = 100~{\rm TeV}$, $g=1$ and $k_S = 0.01$]{
	\includegraphics [width = 7.5cm, clip]{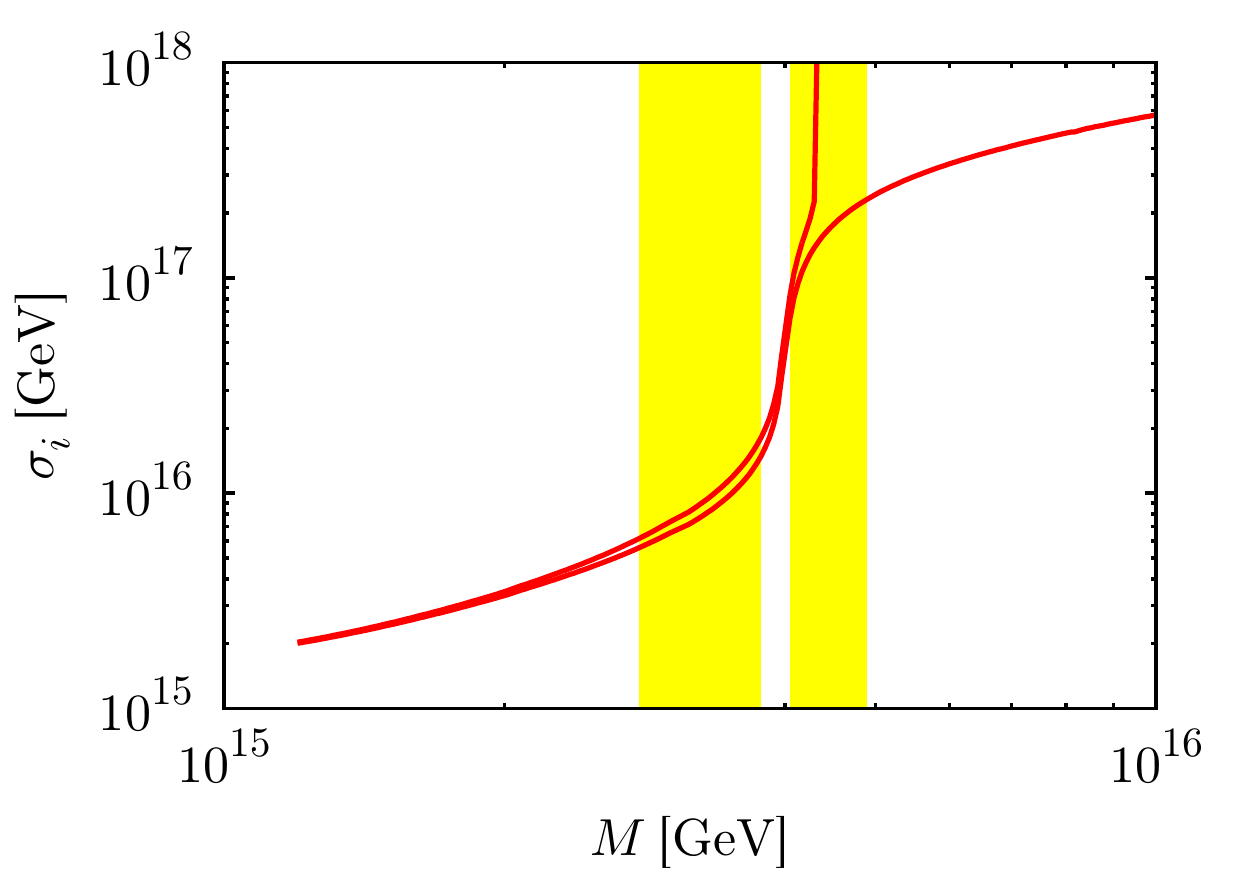}
	\label{Fig4e}
}
\subfigure[$m_{3/2} = 1000~{\rm TeV}$, $g=1$ and $k_S = 0.01$]{
	\includegraphics [width = 7.5cm, clip]{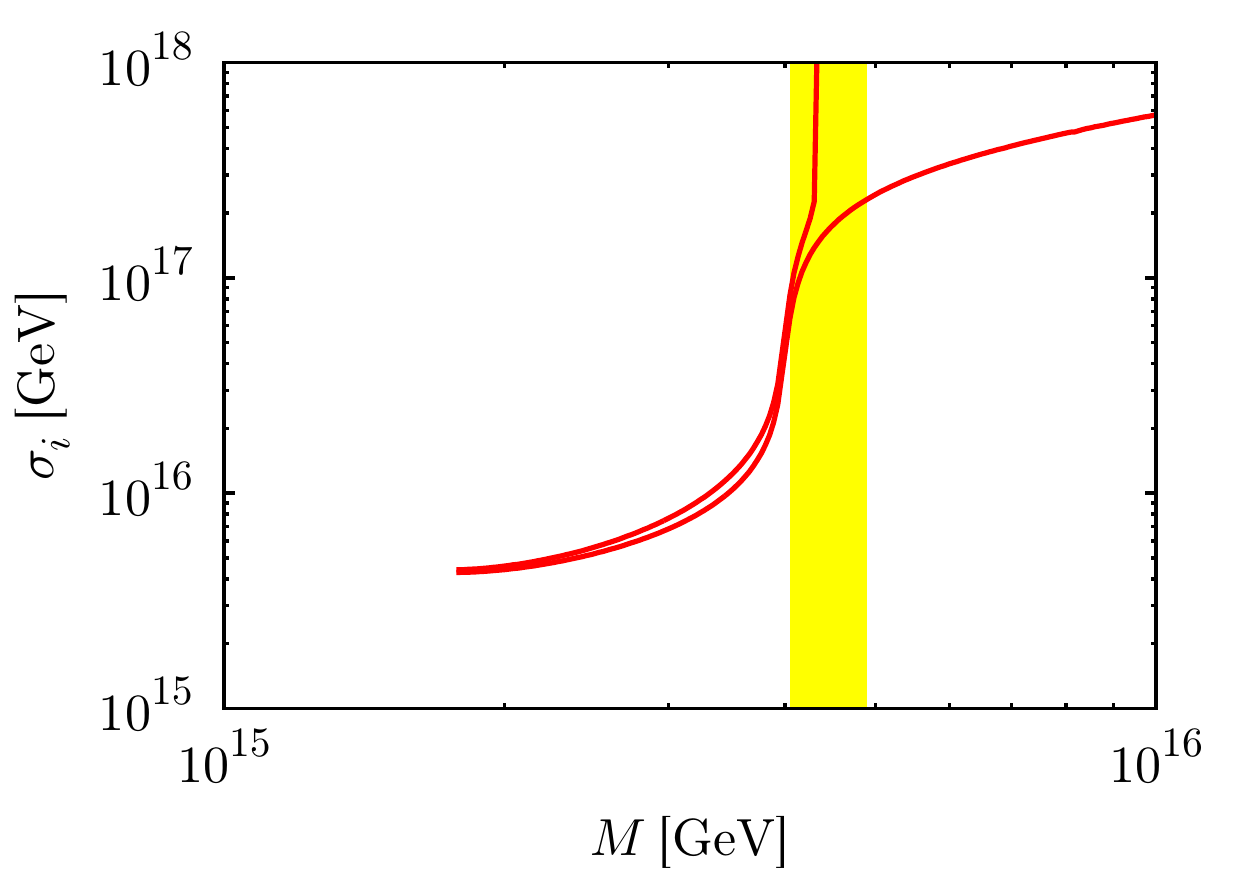}
	\label{Fig4f}
}

\caption{
	Constraints on the initial value of the inflaton are shown.
	We have imposed the WMAP normalization and the allowed regions for the initial values are inside the red contours.
	We have taken $n=4$, $m_{3/2} = 100~{\rm TeV}$ (Fig.~\ref{Fig4a}, \ref{Fig4c}, \ref{Fig4e}), $m_{3/2} = 1000~{\rm TeV}$ (Fig.~\ref{Fig4b}, \ref{Fig4d}, \ref{Fig4f}), 
	$g=1$ (Fig.~\ref{Fig4a}, \ref{Fig4b}, \ref{Fig4e}, \ref{Fig4f}), $g=0.01$ (Fig.~\ref{Fig4c}, \ref{Fig4d}), 
	$k_S=0$ (Fig.~\ref{Fig4a}~-~\ref{Fig4d}) and $k_S = 0.01$ (Fig.~\ref{Fig4c}, \ref{Fig4d}).
	The yellow shaded regions correspond to $0.95 < n_s < 0.98$.
}
\label{Fig4}
\end{figure}

First, let us see the constraint on initial value for the radial component.
In the traditional model the existence of the non-minimal K\"ahler potential, 
the $k_S$ term in (\ref{K_H}) with $k_S > 0$, gives the local minimum 
for the inflaton potential, 
so the initial position of the radial component of the inflaton must be chosen 
not to be trapped at the local minimum.
A similar problem exists in the modified model even if we adopt $k_S = 0$ 
because of the $-v^4\sigma^2/2$ term in the inflaton potential (\ref{V_sigma}).
In order not to be trapped at the local minimum, the initial value of the inflaton 
must be placed smaller than the local maximum, so we require 
$\sigma_i < \sigma_{\rm max}$, where 
\be
	\sigma_{\rm max} \simeq \frac{\kappa}{2\pi\sqrt{2k_S + 2v^4/\kappa^2M^4}}.
\ee
Note that the local minimum and maximum do not arise for $V'(\sigma_*) > 0$ 
where $\sigma_*$ is defined via $V''(\sigma_*) = 0$.
Furthermore, the initial value of the inflaton must be chosen so that 
the inflation last at least 50 e-foldings to solve the horizon problem, hence we impose $\sigma_i > \sigma(N_e)$ with $N_e = 50$, 
where $\sigma_i$ denotes the initial value of the inflation and 
$\sigma(N_e)$ denotes the field value corresponding to the e-folding number $N_e$. 
Fig.~\ref{Fig4} shows the allowed initial values for the inflaton, 
which are inside the red contours.
We have imposed the WMAP normalization on $M$ and $\kappa$ and 
we have taken $n=4$, $m_{3/2} = 100~{\rm TeV}$ (Fig.~\ref{Fig4a}, \ref{Fig4c}, \ref{Fig4e}), $m_{3/2} = 1000~{\rm TeV}$ (Fig.~\ref{Fig4b}, \ref{Fig4d}, \ref{Fig4f}), 
$g=1$ (Fig.~\ref{Fig4a}, \ref{Fig4b}, \ref{Fig4e}, \ref{Fig4f}), $g=0.01$ (Fig.~\ref{Fig4c}, \ref{Fig4d}), $k_S=0$ (Fig.~\ref{Fig4a}~-~\ref{Fig4d}) and $k_S = 0.01$ (Fig.~\ref{Fig4c}, \ref{Fig4d}).
In the yellow shaded region, $0.95 < n_s < 0.98$ is realised.
We have found that, for the minimal K\"ahler potential, $m_{3/2} \lesssim 100~{\rm TeV}$ is disfavored because the extremely fine tuning for the initial value is necessary to reproduce $n_s < 0.98$.
Thus the large gravitino mass is more favored  from the initial value problem.

\begin{figure}[tp]
\centering
\subfigure[$k_S = 0$]{
	\includegraphics [width = 7.5cm, clip]{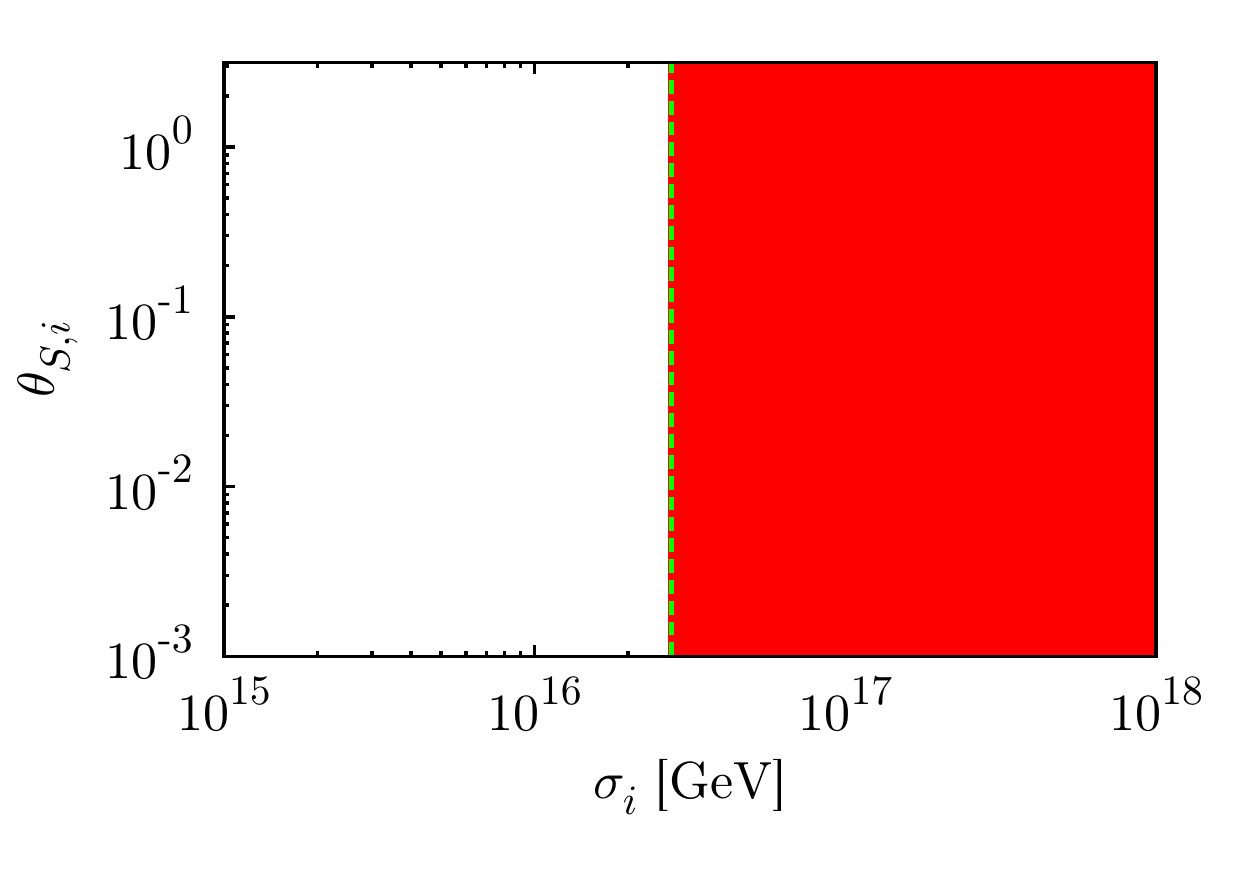}
	\label{Fig3a}
}
\subfigure[$k_S = 0.01$]{
	\includegraphics [width = 7.5cm, clip]{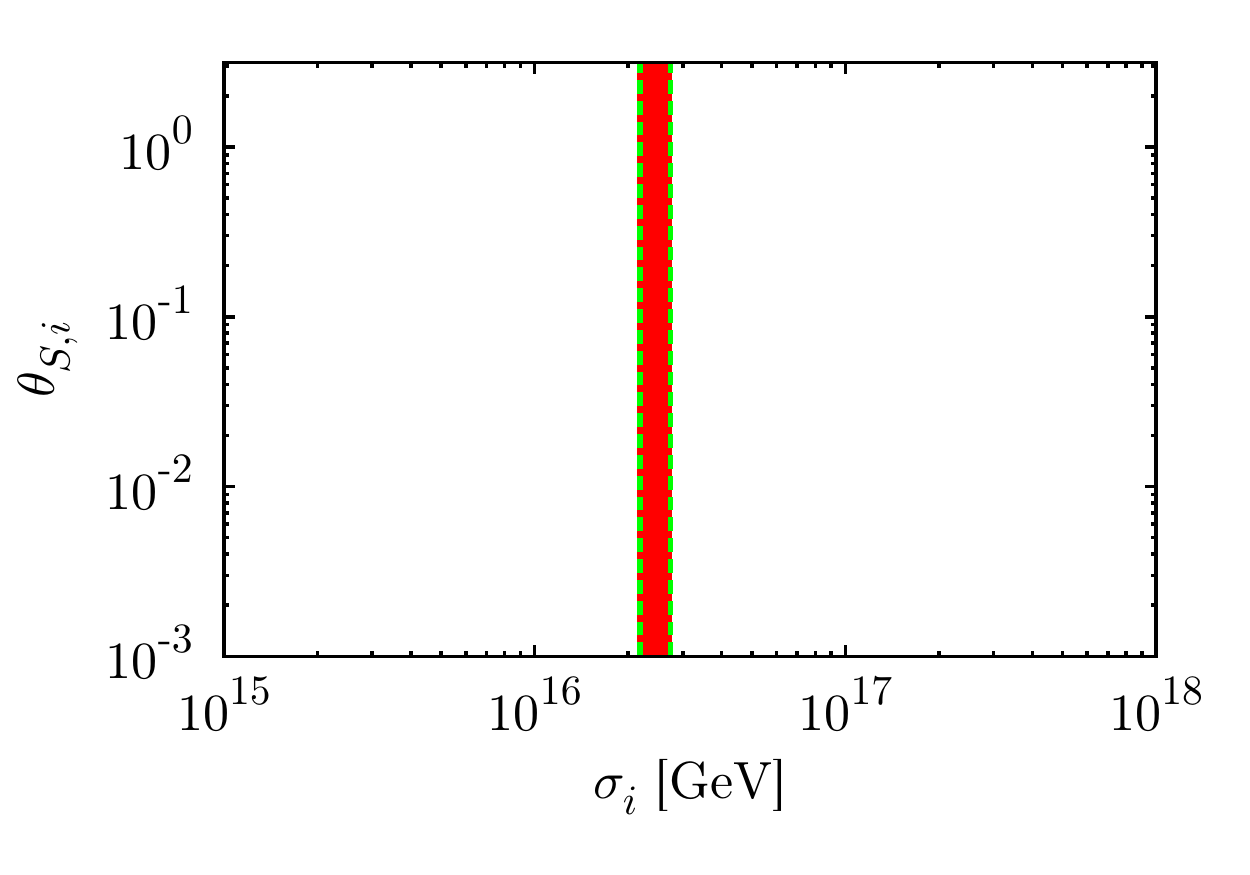}
	\label{Fig3b}
}
\subfigure[$k_S = 0$]{
	\includegraphics [width = 7.5cm, clip]{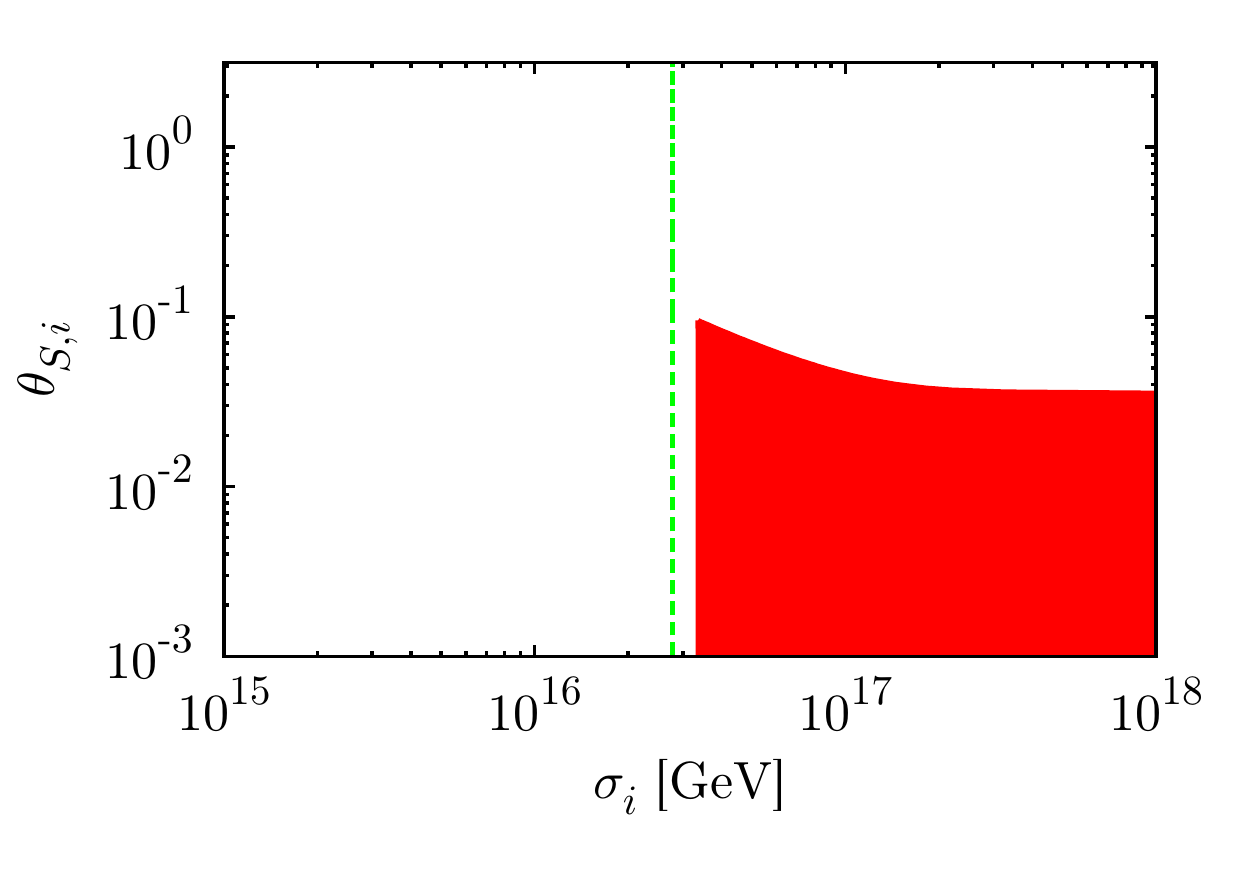}
	\label{Fig3c}
}
\subfigure[$k_S = 0.01$]{
	\includegraphics [width = 7.5cm, clip]{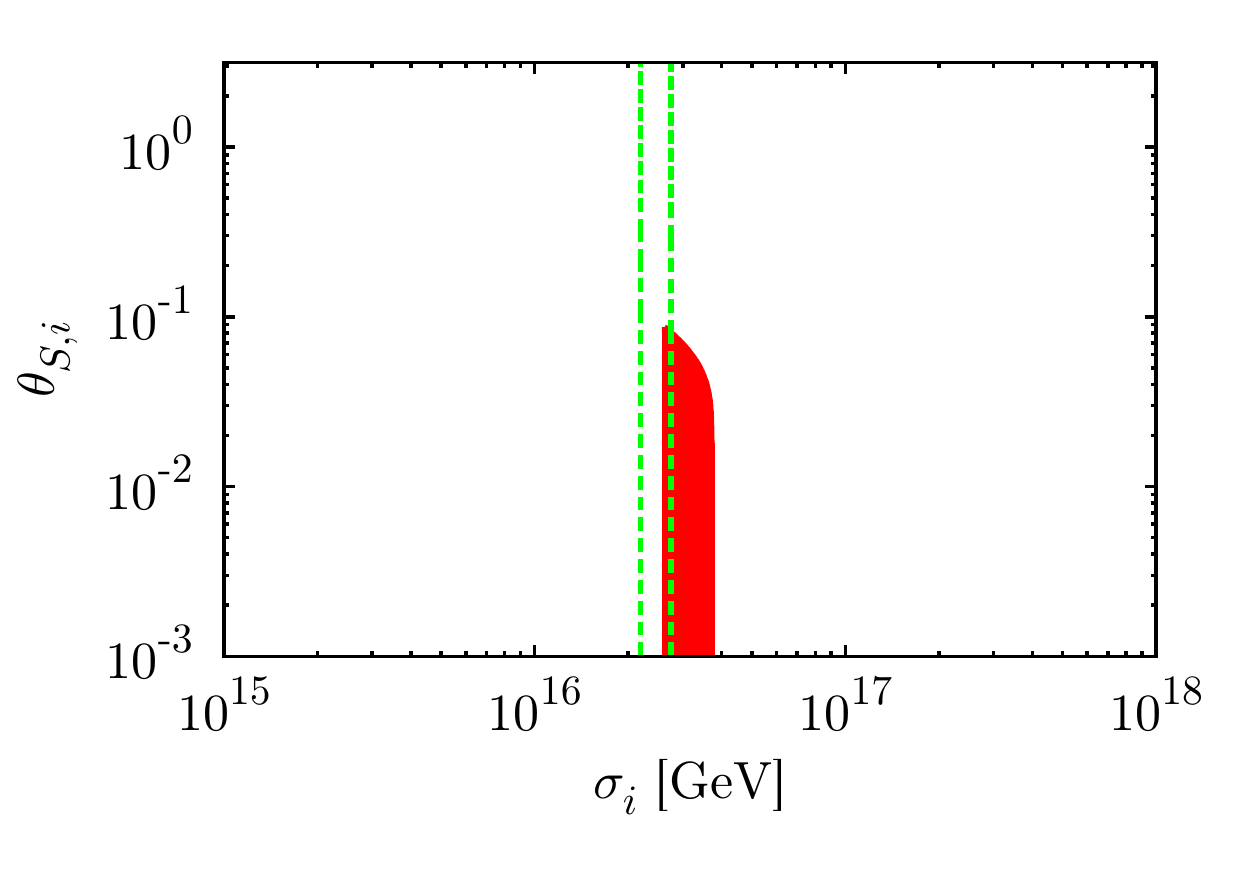}
	\label{Fig3d}
}
\caption{
	We show the range of the allowed initial values of the inflaton as the red regions.
	Fig.~\ref{Fig3a} and \ref{Fig3b} corresponds to our present model and Fig.~\ref{Fig3c} and \ref{Fig3d} corresponds to the traditional SUSY hybrid inflation model with the constant superpotential $W_0$.
	We have taken $M=10^{15}~{\rm GeV}$, $\kappa = 0.01$ and $m_{3/2} = 100~{\rm TeV}$ in all figures and 
	$n = 4$, $g=1$ (Fig.~\ref{Fig3a}, \ref{Fig3b}), $k_S = 0$ (Fig.~\ref{Fig3a}, \ref{Fig3c}), $k_S = 0.01$ (Fig.~\ref{Fig3b}, \ref{Fig3d}).
	We have imposed that the inflation occurs at least for 50 e-foldings, which places the lower bound on the initial value of the inflaton.
	The dashed green lines represent the lower or upper bound for $m_{3/2} = 0$.
}
\label{Fig3}
\end{figure}

On the other hand, in the traditional model, due to constant term $W_0 = m_{3/2}M_P^2$ in the superpotential, the linear term $\sqrt{2} \kappa M^2 m_{3/2} \sigma \cos\theta_S$ arises in the potential for the inflaton, 
where $\theta_S$ denotes the phase component of the inflaton $S$.
For $\cos \theta_S < 0$, the local minimum can be induced, which may trap the inflaton and inflation cannot end.
Hence, the initial phase of the inflaton, $\theta_{S,i}$, must be placed near $\theta_S \sim 0$ so that the angular motion is suppressed and the local minimum does not arise.
This constraint becomes more stringent for larger gravitino mass as shown below.
In the modified model, however, the potential of the inflaton is independent of $\theta_S$, so the initial phase does not affect the dynamics of the inflaton and it can be chosen freely.
To contrast the modified model with the traditional one, we illustrate the allowed region for the initial value for the inflaton in both models in Fig.~\ref{Fig3}.
Fig.~\ref{Fig3a} and \ref{Fig3b} correspond to the present modified model 
and Fig.~\ref{Fig3c} and \ref{Fig3d} correspond to the traditional one with a constant superpotential $W_0$.
In these figures, we have not imposed the WMAP normalization and we have taken $m_{3/2} = 100~{\rm TeV}$, $M = 10^{15}~{\rm GeV}$ and $\kappa = 0.01$ in all figures.
Fig.~\ref{Fig3a} and \ref{Fig3c} correspond to the minimal K\"ahler potential and Fig.~\ref{Fig3b} and \ref{Fig3d} correspond to the non-minimal K\"ahler potential with $k_S=0.01$.
We again emphasize that, compared with the traditional case in which $\theta_{S,i} \lesssim 0.1$ is necessary, the initial phase is not constrained in our present model.

%%%%%%%%%%%%%%%%%%%%%%% FIGURE  %%%%%%%%%%%%%%%%%%%%%%%%%%%%%%%%%%%%%%
%%%%%%%%%%%%%%%%%%%%%%%%%%%%%%%%%%%%%%%%%%%%%%%%%%%%%%%%%%%%%%%%%%%%%%

%%%%%%%%%%%%%%%%%%%%%%% FIGURE  %%%%%%%%%%%%%%%%%%%%%%%%%%%%%%%%%%%%%%
%%%%%%%%%%%%%%%%%%%%%%%%%%%%%%%%%%%%%%%%%%%%%%%%%%%%%%%%%%%%%%%%%%%%%%

%%%%%%%%%%%%%%%%%%%%%%%%%%%%%%%%%%%%%%%%%%%%%%%%%%%%%%%%%%%%%%%%%%%%%%
\subsection{Gravitino problem} \label{gravitino_prob}
%%%%%%%%%%%%%%%%%%%%%%%%%%%%%%%%%%%%%%%%%%%%%%%%%%%%%%%%%%%%%%%%%%%%%%

Gravitinos are copiously produced from the thermal bath at the reheating~\cite{Bolz:2000fu} 
and also non-thermally by the decay of inflaton~\cite{Kawasaki:2006gs}. 
They are often problematic in cosmology.
The abundance of thermally produced gravitinos is sensitive to the reheating temperature
and that of  non-thermally produced ones depends on the inflaton mass and VEV.
The hybrid inflation models often conflict with the gravitino problem since the reheating temperature is expected to be rather high
and the inflaton mass and VEV are also large.

Fortunately, in the modified model, the reheating is induced by the decay of $\Phi$, not the inflaton of hybrid inflation.
This is because the mass of $\Phi$ is much lighter than that of $S$ or $\Psi$, hence we expect that 
$\Phi$ dominates the Universe and decays well after the decay of $S$.
Therefore we can naturally assume that the reheating temperature is determined by the decay of $\Phi$.
All relics from the decay of $S$ are diluted away and become negligible after the decay of $\Phi$.
The abundance of the thermally produced gravitino is estimated as~\cite{Bolz:2000fu}
\be
	Y_{3/2}^\mathrm{(TP)} \simeq 2 \times 10^{-12} \bigg( 1 +\frac{m_{\tilde g}^2}{3 m_{3/2}^2} \bigg) \bigg(\frac{T_R}{10^{10}\,\mathrm{GeV}} \bigg)
\ee
where $T_R$ denotes the reheating temperature and $m_{\tilde g}$ is the gluino mass evaluated at $T=T_R$.
Gravitinos are also produced non-thermally from the direct decay of $\Phi$~\cite{Kawasaki:2006gs}.
The decay is induced by, e.g., the following non-renormalizable operator in the K\"ahler potential : $K \sim |\Phi|^2 zz +{\rm h.c.}$
where $z$ denotes the SUSY breaking field. The resulting gravitino abundance is estimated as
\be
	Y_{3/2}^{\rm (NTP)} \simeq 7 \times 10^{-14} \bigg( \frac{10^7\,{\rm GeV}}{T_R} \bigg)
	\bigg( \frac{m_{\phi}}{10^{10}\,{\rm GeV}} \bigg)^2 \bigg(\frac{\Phi_{0}}{10^{16}\,{\rm GeV}} \bigg)^2.
\ee
In the dynamical SUSY breaking where $z$ is charged under some symmetry, 
such operators are forbidden.
The $\Phi$ decay into hidden sector hadrons are also forbidden since $\Phi$ 
is lighter than the SUSY breaking scale.
Then the decay rate into gravitinos are suppressed by the factor $\sim (m_z/m_\phi)^4$ 
where $m_z$ is the mass of SUSY breaking field and constraint from nonthermal gravitino 
production is significantly relaxed~\cite{Nakayama:2012hy}.

\begin{figure}[tp]
\centering
\subfigure[$n = 4$]{
	\includegraphics [width = 7.5cm, clip]{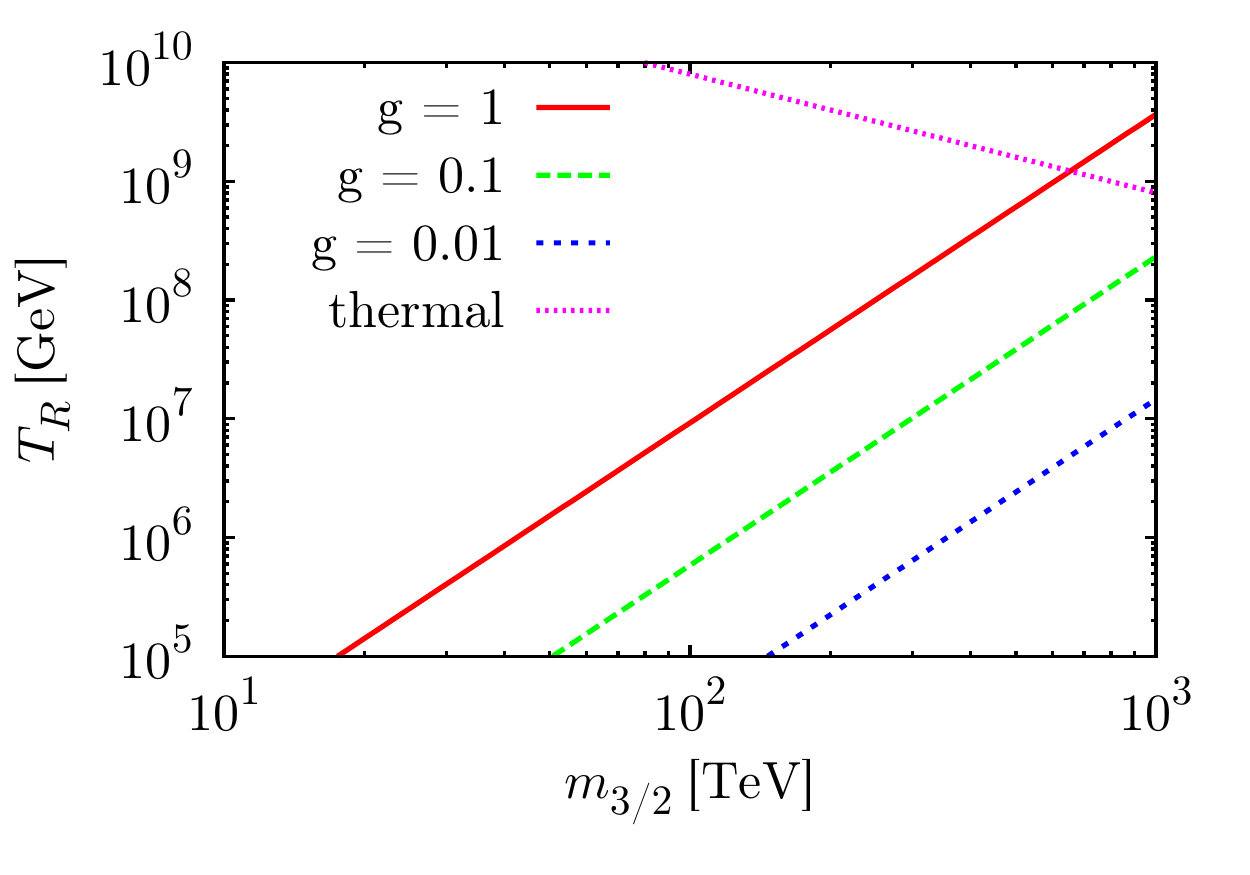}
	\label{Fig5a}
}
\subfigure[$n = 6$]{
	\includegraphics [width = 7.5cm, clip]{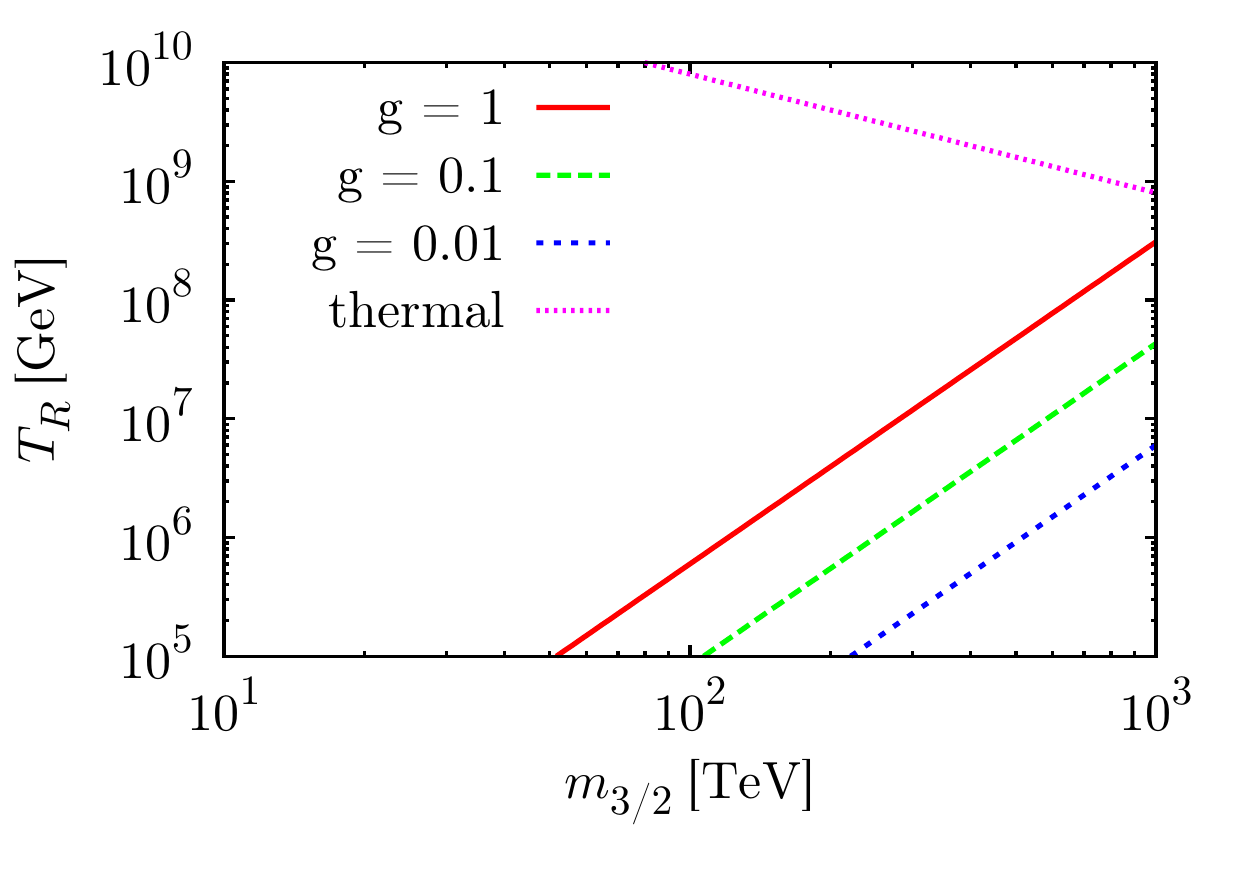}
	\label{Fig5b}
}
\caption{
	Constraints on $T_R$~-~$m_{3/2}$ plane from the gravitino abundance are shown.
	The solid red lines, dashed green lines and dotted blue lines represent lower bounds on reheating temperature 
	from the non-thermally produced gravitinos 
	and small-dotted magenta line represents upper bound from the thermally produced one.
	We have taken $n=4$ (Fig.~\ref{Fig5a}), $n=6$ (Fig.~\ref{Fig5b}) and $g=1$ (solid red lines), $g=0.1$ (dashed green line) 
	and $g = 0.01$ (dotted blue lines). 
	We have assumed the anomaly-mediated SUSY breaking model in which $m_{\rm LSP} \simeq m_{3/2}/400$ is satisfied.
}
\label{Fig5}
\end{figure}

For $m_{3/2} \gg 10\,{\rm TeV}$, gravitinos eventually decay into 
the lightest SUSY particles (LSPs) 
well before the big bang nucleosynthesis (BBN).
In order for such LSPs not to overclose the Universe,  the constraint on the gravitino 
abundance is given by
\be
	m_{3/2} Y_{3/2} \lesssim 4 \times 10^{-10}~{\rm GeV} 
	\bigg( \frac{m_{3/2}}{m_{\rm LSP}} \bigg).
	\label{constraint_gravitino}
\ee
Assuming the anomaly-mediated SUSY breaking model~\cite{Randall:1998uk,Giudice:1998xp} 
or pure gravity-mediation model~\cite{Ibe:2011aa}, the LSP is the neutral Wino 
whose mass is given by $M_2 = (g_2^2/16\pi^2) m_{3/2} \simeq m_{3/2} / 400$ 
where $g_2$ is the weak coupling constant.
Using this relation and (\ref{constraint_gravitino}), the reheating temperature 
is constrained as shown in Fig.~\ref{Fig5}.
The allowed parameters are below the small-dotted magenta lines 
and above the solid red lines for $g=1$ (above the dashed green lines for $g=0.1$ 
or dotted blue lines for $g=0.01$).
The situation is much better than the case of traditional hybrid inflation model 
where the nonthermal gravitino production poses a stringent 
constraint~\cite{Nakayama:2010xf}.
Note that, by choosing the parameters properly, the observed dark matter abundance 
can be explained by the non-thermally-produced Wino-LSP from the gravitino decay 
as shown in~\cite{Ibe:2006ck} 
and it may be detected by experiments such as LHC and 
so on~\cite{Moroi:2011ab,Bhattacherjee:2012ed,Hall:2012zp}.

%%%%%%%%%%%%%%%%%%%%%%% FIGURE  %%%%%%%%%%%%%%%%%%%%%%%%%%%%%%%%%%%%%%
%%%%%%%%%%%%%%%%%%%%%%%%%%%%%%%%%%%%%%%%%%%%%%%%%%%%%%%%%%%%%%%%%%%%%%

%%%%%%%%%%%%%%%%%%%%%%%%%%%%%%%%%%%%
\section{Conclusion}\label{conc}
%%%%%%%%%%%%%%%%%%%%%%%%%%%%%%%%%%%%

We have revisited the SUSY hybrid inflation model focusing on the large gravitino mass case,
motivated by the recent LHC results showing the Higgs mass around $m_h \simeq 125~{\rm GeV}$~\cite{:2012gk}. 
Instead of the constant term in the superpotential which is required to cancel the vacuum energy, 
we have replaced it with a dynamical field, which effectively becomes a constant term at the present vacuum.
We have shown that the allowed parameters are altered from the traditional case and, in particular, the relatively large gravitino mass  $m_{3/2} \sim 100-1000~{\rm TeV}$ is consistent with hybrid inflation in our model.
The constraint from the cosmic string is also relaxed.
Furthermore, the observed spectral index can also be reproduced without invoking the non-minimal K\"ahler potential. 
Although the initial value of the radial component of the inflaton must be tuned so as not to be trapped at the local minimum, 
the initial phase component of the inflaton is not constrained, 
hence the initial value problem becomes milder compared with the traditional model.
%The cosmological gravitino problem was also considered.
%Because the reheating does not occur at the decay of $S$ but occurs at the decay of $\Phi$, the reheating temperature is reduced significantly compared with the traditional model and 
Since the reheating is induced by the dynamical field $\Phi$, not the inflaton,
the cosmological gravitino problem can easily avoided in our model.

%%%%%%%%%%%%%%%%%%%%%%%%%%%%%%%%%%%%
\section*{Acknowledgment}
%%%%%%%%%%%%%%%%%%%%%%%%%%%%%%%%%%%%

This work is supported by Grant-in-Aid for Scientific research from
the Ministry of Education, Science, Sports, and Culture (MEXT), Japan,
No.\ 22540267 (M.K.), No.\ 21111006 (M.K. and K.N.), No.\ 22244030 (K.N.) and also 
by World Premier International Research Center
Initiative (WPI Initiative), MEXT, Japan. 
N.K. is supported by the Japan Society for the Promotion of Science (JSPS).

%\input{bib}

%%%%%%%%%%%%%%%%%%%%%%%%%%%%%%%%%%%%%%%%%%%%

%%%%%%%%%%%%%%%%%%%%%%%%%%%%%%%%%%%%%%%%%%%%

\end{document}